\documentclass[a4paper,fleqn]{cas-sc}

\usepackage[numbers]{natbib}
\usepackage{indentfirst} 

\usepackage{algorithm}
\usepackage{algorithmic}

\usepackage{graphicx}

\def\tsc#1{\csdef{#1}{\textsc{\lowercase{#1}}\xspace}}
\tsc{WGM}
\tsc{QE}
\tsc{EP}
\tsc{PMS}
\tsc{BEC}
\tsc{DE}

\newcommand{\fanxy}[1]{{#1}}

\begin{document}

\let\WriteBookmarks\relax
\def\floatpagepagefraction{1}
\def\textpagefraction{.001}


\shortauthors{fanhang et~al.}

\title [mode = title]{Privacy Preserving Ultra-Short-term Wind Power Prediction Based on Secure Multi Party Computation}                      


%
\author[1]{Hang Fan}



\ead{fanhang123456@163.com}


\credit{Conceptualization of this study, Methodology, Software}

\address[1]{organization={PBC School of Finance},
    addressline={Tsinghua University}, 
    city={Beijing},
    postcode={100084}, 
    country={China}}

\author[2]{Xiaoyu Fan}
\address[2]{organization={Institute for Interdisciplinary Information Sciences},
    addressline={Tsinghua University}, 
    city={Beijing},
    postcode={100084}, 
    country={China}}
    
\author[1]{Tianyi Hao}

\credit{Data curation, Writing - Original draft preparation}

\author%
[3]
{Wei Wei}

\address[3]{organization={Department of Electrical Engineering},
    addressline={Tsinghua University}, 
    city={Beijing},
    postcode={100084}, 
    country={Beijing}}

\author%
[4]
{Kun Chen}

\author%
[4]
{Guosai Wang}

\address[4]{organization={Tsingjiao Information Technology Co. Ltd.},
    addressline={Tsinghua Science and Technology Park}, 
    city={Beijing},
    postcode={100084}, 
    country={Beijing}}

\author%
[5]
{Xiaofeng Jia}

\address[5]{organization={Data Management Department},
    addressline={Beijing Big Data Center}, 
    city={Beijing},
    postcode={100044}, 
    country={Beijing}}

\author%
[6]
{Yidong Li}

\address[6]{organization={School of Computer and Information Technology},
    addressline={Beijing Jiaotong University}, 
    city={Beijing},
    postcode={100025}, 
    country={Beijing}}

\author%
[2]
{Wei Xu}
\cormark[1]
\ead{weixu@mail.tsinghua.edu.cn}
\cortext[cor1]{Corresponding author}


\begin{abstract}
Mining the spatial and temporal correlation of wind farm output data is beneficial for enhancing the precision of ultra-short-term wind power prediction. However, if the wind farms are owned by separate entities, they may be reluctant to share their data directly due to privacy concerns as well as business management regulation policies. Although cryptographic approaches have been designed to protect privacy in the process of data sharing, it is still a challenging problem to encrypt the original data while extracting the nonlinear relationship among multiple wind farms in the machine learning process. 
This paper presents \textbf{pwXGBoost}, a technique based on the machine learning tree model and secure multi-party computation (SMPC) that can successfully extract complicated relationships while preserving data privacy. 
A maximum mean discrepancy (MMD) based scheme is proposed to effectively choose adjacent candidate wind farms to participate in the collaborative model training, therefore improving the accuracy and reducing the burden of data acquisition. 
The proposed method was evaluated on real world data collected from a cluster of wind farms in Inner Mongolia, China, demonstrating that it is capable of achieving considerable efficiency and performance improvements while preserving privacy.
\end{abstract}


\begin{highlights}
\item We develop a vertical privacy preserving XGBoost prediction algorithm based on the secret sharing protocol in the pwXGBoost model.
\item We design a criterion to select the suitable participant wind farm in the pwXGBoost model. 
\item We test the wind farms in the field data from the wind farm cluster in the Inner Mongolian.
\end{highlights}


\maketitle

\section{Introduction} \label{sec:intro}
\subsection{Background and Motivation}
{T}{he} large-scale exploitation of renewable energy sources, such as wind power, has brought a large amount of clean energy and reduced the CO2 emissions. However, the uncertainty and randomness of wind power pose serious challenges to the operation of the power system~\cite{he2016cooperation}. On the electricity spot trading market, the bidding strategies of wind farms heavily depend on the wind power predictions and severe deviations in bids are penalized. The economic losses caused by inaccurate wind power forecasts can reach 10\% of the wind farms' electricity sales \cite{fabbri2005assessment}. \par
Due to the desire for high-quality wind power prediction, there have been tremendous studies in related fields. At present, the core concept of ultra-short-term power prediction is to mine a variety of data such as historical power generation data and \textit{Numerical Weather Prediction (NWP)} data of local wind sites through artificial intelligence and statistical learning methods, so as to develop high-precision prediction models. 
At the same time, wind farms are usually located in close proximity to each other, there is a strong spatial and temporal correlation pattern among the sites. Utilizing this correlation pattern can considerably enhance the power prediction accuracy of wind farms.
As a result, more studies have been conducted in recent years by assuming that the data of all wind farms in the cluster can be obtained. Through combining graph machine learning and other nonlinear methods to extract the spatial and temporal correlation among neighboring wind farms, the prediction accuracy can be effectively improved \citep{tastu2013probabilistic,fan2020m2gsnet}.  
However, historical wind power data is often owned by different companies. The long-time historical wind power can reflect the production and operation status of the wind farms in the electricity market and is confidential to each other. 
The NWP data is purchased by wind farms at high cost, thus they are hesitant to share with others. Therefore, the direct sharing of wind power data may be restricted by data management policies due to the privacy and security of the data. 
How to use data from neighboring wind farms without compromising privacy is considered as the last mile and the most difficult part of the application of spatial and temporal correlation methods in practice.\par

\subsection{Previous Study and Literature Review}
Some research is brought out to predict the wind power while preserving the original data. For this problem, article \cite{gonccalves2021critical} classify the solutions into three classes, namely \textit{data transformation \fanxy{method, }} decomposition method and secure multi-party computation method. According to the definition, the data transformation method normally refer to adding some random noise to the original data before \fanxy{the} fitting process to protect the privacy which is called differential privacy \cite{dwork2008differential}. Although the differential privacy is successful to protect the privacy in the picture recognition area, it is not suitable for the wind power prediction \cite{abadi2016deep}. The picture recognition is a classification problem while the wind power prediction is a regression problem which is more sensitive to the input data. Any disturb to the wind power data can lead to a decrease of accuracy which is unacceptable for the power market and the wind farm owner. \par

\textit{Decomposition method} regards the prediction problem as an optimization problem and decomposes it into several sub-problems and allows each data provider to solve it separately. Carla \cite{goncalves2021privacy} emphasizes the forecast skill improvement due to the spatial and temporal dependencies in the time series and the business competition among wind farms. Therefore, \cite{goncalves2021privacy} formulates a framework which combines the data transformation methods and the \textit{alternating direction method multipliers} (ADMM). In \cite{goncalves2020towards}, a data market even been designed to encourage the wind farms to share their data to improve the prediction accuracy. Han \cite{han2021trading} designed an regression market for wind power forecasting and use the LASSO regulation as the reference for the data pricing. However, wind farm power is highly nonlinear, and the lasso method, as a linear prediction method, is inherently difficult to capture the spatial and temporal correlation among wind farms, so the prediction accuracy in practice is not always satisfactory. In practical wind farm power prediction tasks, multiple nonlinear prediction methods such as machine learning, XGBoost or even neural networks and their combined derivative models are more often used. And as stated in the article \cite{goncalves2021privacy}, privacy methods using ADMM methods for solving cannot be directly extended to nonlinear prediction scenarios. Therefore, there is an urgent demand to explore how to consider data privacy in nonlinear prediction models.\par

\textit{Secure multi-party computation} in article \cite{gonccalves2021critical} is a generalized  privacy preserving computation framework \cite{gonccalves2021critical}. This topic is an active research field in computer science and data mining because it is compatible with non-linear operation \cite{kairouz2021advances}. It calls for the fusion of classical secure multi-party computation \cite{yao1982protocols}, federated learning \cite{yang2019federated} and other classical cryptography theory such as homomorphic encryption \cite{ogburn2013homomorphic}.  \par

Classical secure multi-party computing techniques include secret sharing, oblivious transmission, and \fanxy{garbled} circuit, which are mainly derived from the "millionaire's problem" in 1982 \cite{yao1982protocols}. In 1986, Yao proposed the theory of the \fanxy{garbled} circuit, which became the first general multi-party secure computing scheme \cite{yao1986generate}. After several years of development, the classical secure multi-party computation  consists of multiple cryptography protocols such as garbled circuit \cite{bellare2012foundations}, oblivious transmission \cite{rabin2005exchange} and secret sharing \cite{shamir1979share}. \fanxy{Garbled} circuit is performed by constructing a circuit and obfuscating the signals on the circuit, while secret sharing is performed by splitting the secret data into multiple slices and performing computation on the slices. Because secret sharing protocol is more friendly for the computation, most advanced privacy preserving computation platform adopt this protocol \cite{li2019privpy,keller2020mp}, and it quickly becomes a popular method in recent studies. For example, article \cite{tian2021fully} uses the secret share to realize the fully privacy preserving distributed optimization of power system.\par 

Federated learning is a distributed machine learning method proposed by Google in 2016 \cite{li2020federated} that enables multiple mutually untrusted training data providers to collaboratively train machine learning models by exchanging intermediate computational results such as gradients or parameters without exchanging raw data. According to the different data distribution among participants, federation learning is generally classified into three types: horizontal federated learning \cite{kairouz2021advances}, vertical federated learning \cite{kairouz2021advances} and federated transfer learning \cite{liu2020secure}. Horizontal federation is mainly used for sample federation between two parties with the same or similar business model, and there is a lot of feature overlap in the data of each party, but less overlap in the number of users. Longitudinal federation is mainly used for feature federation between two parties with different business modes but the same or similar users, with less feature overlap but more user overlap. Federated migration is mainly used for forward learning between two parties with less intersection of industry and users, and there is less overlap of features and users in the data of all parties.
There are many scenarios that the federated learning is used to protect the privacy. \cite{toubeau2022privacy} used the federated learning for the voltage prediction in the local energy community. In \cite{wang2023federated}, the federated fuzzy k-means is used to analyze the smart grid meter data. Federated learning can also be used with the reinforcement learning. In \cite{qiu2023federated}, a federated reinforcement learning method is designed for the peer-to-peer energy trading and the carbon allowance trading. Article \cite{li2023wind} uses the horizontal federated reinforcement learning to predict the wind power which can leverage the wind farms in a cluster. However, it can not extract the wind farm spatialtemporal relationship which is implied by the wind power data at the same time. In 2019, it was demonstrated that the gradients or parameters exchanged during federation learning can be used to infer or even recover the original data information \cite{sun2019can}, currently, the exchange process usually requires cryptography-based techniques (e.g., MPC) or homomorphic cryptography to avoid these risks \cite{acar2018survey}. On the other hand, the performance of the federated learning will decrease if the data distribution of the participants are non-iid such as the feature distribution skew, label distribution skew and quantity skew \cite{li2021federated}. Take the wind power prediction case for example, if the wind data of the participants do not follow similar distributions or appear to be non-iid, it is more difficult to identify the spatial temporal correlation patterns.\par

Homomorphic encryption is a classical encryption method to protect data privacy by directly encrypting the plaintext, performing various operations under the ciphertext, and finally obtaining the resulting ciphertext. Homomorphic encryption can be classified into Fully Homomorphic, Somewhat Homomorphic, and Partially Homomorphic \cite{fontaine2007survey,acar2018survey} depending on the degree of support for an unlimited number of arbitrary homomorphic operations. Homomorphic encryption allows arbitrary computation of the ciphertext without decryption, but its performance is too slow to become practical. According to the latest Fully Homomorphic computation benchmark~\cite{gouert2022new} in 2022, the homomorphic computation is orders of magnitude slower than plaintext. The limited computational speed constrains its practical application \cite{wang2012accelerating}.\par

\subsection{Contribution and Paper Organization}

Although there are some works aim to preserve the privacy in the prediction process, there are still two main problems. The first problem is the current privacy preserving method can not fully utilize the spatial and temporal correlation while safely preserving the data privacy. Although some researchers use the federated learning method such as the FedAVG to fusion the gradient of the neural network, it follows the horizontal data fusion method. 
It is more similar to the transfer learning which is suitable when the wind power data is sufficient and horizontally partitioned among wind farms rather than the case that the wind farm can boost his own prediction accuracy by utilizing the spatial and temporal relationship with others.
Moreover, the classic federated learning method is not safe enough for the collaborative modeling and prediction \cite{sun2019can}. Current privacy preserving prediction method which can extract the spatial and temporal relationship is based on Lasso-var and it is a linear method \cite{goncalves2021privacy}. But the spatial and temporal relationship is highly nonlinear, and the feature extraction ability of linear method is limited. Using homomorphic encryption and other full ciphertext computing methods can solve the nonlinear problem in the extraction of spatial and temporal correlation, but the expensive computation cost makes it impractical \cite{acar2018survey}. The second problem is the participant selection. If the data of each wind farm exhibits significant difference in their distributions, the non-iid feature will effect the performance of prediction. Besides, if the number of wind farms in the collaborated power prediction is extremely great, the communication cost will compromise the timeliness of the ultra-short-term prediction model. If insufficient wind farms participate in the collaborative model training, spatial and temporal correlation will not be utilized to its full potential. Therefore, we designed a method named \textbf{pwXGBoost} based on vertical data fusion strategy and the secret sharing protocol which is scalable to extract the nonlinear spatial and temporal correlation pattern of several wind farms. In the pwXGBoost model, we also borrowed ideas from personalized federation learning \cite{tan2022towards} to screen participants for the collaborative modeling task.\par

The contribution of this paper is three-fold:\par
(1) We develop a vertical privacy preserving XGBoost prediction algorithm based on the secret sharing protocol in the pwXGBoost model. It has the following advantages. First, it is scalable to the nonlinear data and complex modeling of the spatial and temporal correlation compared to the renowned Lasso-var method. Second, it can realize a lossless and secure computation of XGBoost. It is more precise than the data transformation methods and more secure than the conventional federated learning. \par
(2) We design a criterion to select the suitable participant wind farm in the pwXGBoost model. In the criterion, during the collaborative training process the maximum mean discrepancy index is adopted to assess the similarity of the wind farm data distribution and it can select the wind farm which is most useful for the spatial and temporal correlation extraction. \par
(3) We test the wind farms in the field data from the wind farm cluster in the Inner Mongolian. The data of some nearby wind farms in the cluster are combined to predict the wind power of the target wind farm. The experiment results show that the proposed pwXGBoost method is superior than all the baseline methods which only uses local data or a linear model. The prediction time is also acceptable for the practical application. \par 

The rest of this paper is organized as follows. Section~\ref{sec:prob} will formulate the mathematical model of the privacy preserving ultra-short-term wind power prediction. Section~\ref{sec:windpow} develops the privacy preserving XGBoost model based on secret sharing protocol. Section~\ref{sec:xgboost} describes the implementation process of the privacy preserving ultra-short-term wind power prediction. Section~\ref{sec:secu} analyzes the security of the proposed approach. The privacy preserving prediction algorithm is tested on the wind farm cluster from Inner Mongolian and the effectiveness is validated in Section~\ref{sec:case} Finally. Conclusions are drawn in Section~\ref{sec:conclu}.\par

\section{Preliminary} 
\subsection{Traditional Wind Farm Power Prediction}
Wind power forecast is a classical time series prediction problem which have been extensively studied. For ultra-short-term wind power prediction, traditionally only the data of the local wind farm is used for the modeling.\par

\begin{equation}
 P_{t+H} = f(P_{t-M+1:t},V_{t+1:t+N})
\end{equation}
Where $P_{t-M+1:t}$ is the local historical wind power of the wind farm and $V_{t+1:t+N}$ is the local NWP of the wind farm. Recently, it has been recognized that exploiting the spatial and temporal correlation can improve forecast accuracy \cite{tastu2013probabilistic}. Therefore, the wind power prediction can be modeled as follows: 

\begin{equation}
 P_{t+H}^{i} = f(P_{t-M+1:t}^{1},...,P_{t-M+1:t}^{i},...,P_{t-M+1:t}^{n},...,V_{t+1:t+N}^{1},...V_{t+1:t+N}^{i},...V_{t+1:t+N}^{n})
\end{equation}
Where $P_{t-M+1:t}^{i}\in R^{M\times1}$ is the historical wind power of wind farm $i$ and the step length for the prediction is $M$. $V_{t+1:t+N}^{i}\in R^{N\times k}$ is the matrix of NWP data for wind farm $i$ and the step length for the NWP data is $N$. The variable number of NWP data is $k$. Normally, the next 4 hour wind power are to be predicted and the time interval is 15min, so $H=16$. Function $f$ is the prediction model which can be linear model such as Lasso, neural network or XGBoost model. In the prediction model, although only the wind power of wind farm $i$ needed to be predicted, the historical wind power and NWP data of other nearby wind farms are used. For a wind farm cluster, if the wind farms in this cluster are belong to the same owner, this kind of centralized prediction model is acceptable. However, when the wind farms in the cluster are the assets of different stakeholders, direct data sharing is not appropriate. Therefore, it is necessary to develop the privacy preserving prediction model.

\subsection{A Brief Review of XGBoost}
XGBoost is an ensemble of tree models to boost the performance of a single tree which is very popular in wind power prediction\cite{chen2016xgboost}. For a dataset $\textbf{X} \in R^{M \times N}$ with $M$ samples and $N$ features. XGBoost can predict the $i$-th sample $x_i \in R^{1 \times N}$ by using $T$ regression function as follows:

\begin{equation}
\begin{aligned}
\hat{y}_{i} = \sum_{t=1}^{T}f_{t}(x_{i})
\end{aligned}
\end{equation}

XGBoost is sequentially trained by calculating $\hat{y}_{i}^{(t)} = \hat{y}_{i}^{(t-1)} + f_{t}(x_{i})$, where a new tree $f_{t}(x_{i})$ is used to train the residual of the target and the prediction in the previous iteration. For the given loss function $l$, a second-order Tylor expansion is used to approximate it in $t$-th iteration as follows:

\begin{equation}
\begin{aligned}
\mathcal{L} \approx \sum_{i=1}^{N}[l(y_i,\hat{y}_{i}^{(t-1)})+g_{i}f_{t}(x_{i})+\frac{1}{2}h_{i}f_{t}^{2}(x_{i})] + \Omega(f_t)
\end{aligned}
\end{equation}

\begin{equation}
\begin{aligned}
\Omega(f_t) = \gamma U + \frac{1}{2}\lambda||\omega||^{2}
\end{aligned}
\end{equation}
Where $\hat{y}_{i}^{(t-1)}$ is the current prediction results, $\Omega$ is the regulation term, $U$ is the number of leaves in the tree, $\gamma$ and $\lambda$ are the hyper-parameters to restrict the tree number and weights respectively.
$g_{i}=\partial_{\hat{y}_{i}^{(t-1)}}l(y_i,\hat{y}_{i}^{(t-1)})$ and $h_{i}=\partial^2_{\hat{y}_{i}^{(t-1)}}l(y_i,\hat{y}_{i}^{(t-1)})$ are the first and second order derivative statistics of loss function.
The tree model starts from $s$ single leaf node which includes all samples. Then the node recursively splits the current samples into left and right subsets denoted by $I_{L}$ and $I_{R}$. The loss function after the split is

\begin{equation}
\begin{aligned}
\mathcal{L}_{\mathrm{split}} \approx \frac{1}{2}[\frac{(\sum_{i\in I_{L}}g_{i})^2}{\sum_{i\in I_{L}}h_{i}+\lambda}+\frac{(\sum_{i\in I_{R}}g_{i})^2}{\sum_{i\in I_{R}}h_{i}+\lambda}-\frac{(\sum_{i\in I_{I}}g_{i})^2}{\sum_{i\in I_{I}}h_{i}+\lambda}]-\gamma
\end{aligned}
\end{equation}

Where the best split is the one with the highest $\mathcal{L}_{split}$. The weight $w$ of each leaf is calculated in equation (10)

\begin{equation}
\begin{aligned}
w = -\frac{\sum_{i\in I_{u}}g_{i}}{\sum_{i\in I_{u}}h_{i}+\lambda}
\end{aligned}
\end{equation}
When the depth of the tree reach the highest, the training of XGBoost terminates \cite{chen2016xgboost}. 

\section{Problem Definition} \label{sec:prob}

 Due to the privacy concern, wind farms can hardly share their data directly without any constrains. Because once the wind power data of the wind farms are copied to other places, the risk of data abuse seems inevitable. The privacy-preserving wind farm power prediction allows wind farms to access the data of other adjacent wind farms to train the prediction model jointly without knowing the exact values of those data. In this section, the ultrashort-term wind power forecast problem is formulated as follows:
 

\begin{equation}
\begin{aligned}
 P_{t+H}^{i} = f([P_{t-M+1:t}^{1}],...,P_{t-M+1:t}^{i},...,[P_{t-M+1:t}^{n}],...,[V_{t+1:t+N}^{1}],...V_{t+1:t+N}^{i},...[V_{t+1:t+N}^{n}])
\end{aligned}
\end{equation}
Where $[]$ is the secret share encryption of the variable. Through the secret share encryption, the data and computation is only known by the owner of the data \cite{nishide2007multiparty}. $[P_{t-M+1}^{n}]$ is the vector of the wind power in the secret share cipher text of $P_{t-M+1}^{n}\in R^{M\times1}$ fow wind farm $n$. $[V_{t+1:t+N}^{n}]$ is the matrix in the secret share cipher text of NWP matrix $V_{t+1:t+N}^{n}\in R^{N\times k}$ for wind farm $n$. $M$ is the step length of historical wind power used in the prediction. $N$ is the step length of NWP data used in the prediction. Prediction function $f$ uses the cipher text of the nearby wind farms to predict their own wind power in the next few hours. \par

\begin{figure*}[htbp]
    \centering
    \includegraphics[width=0.8\textwidth]{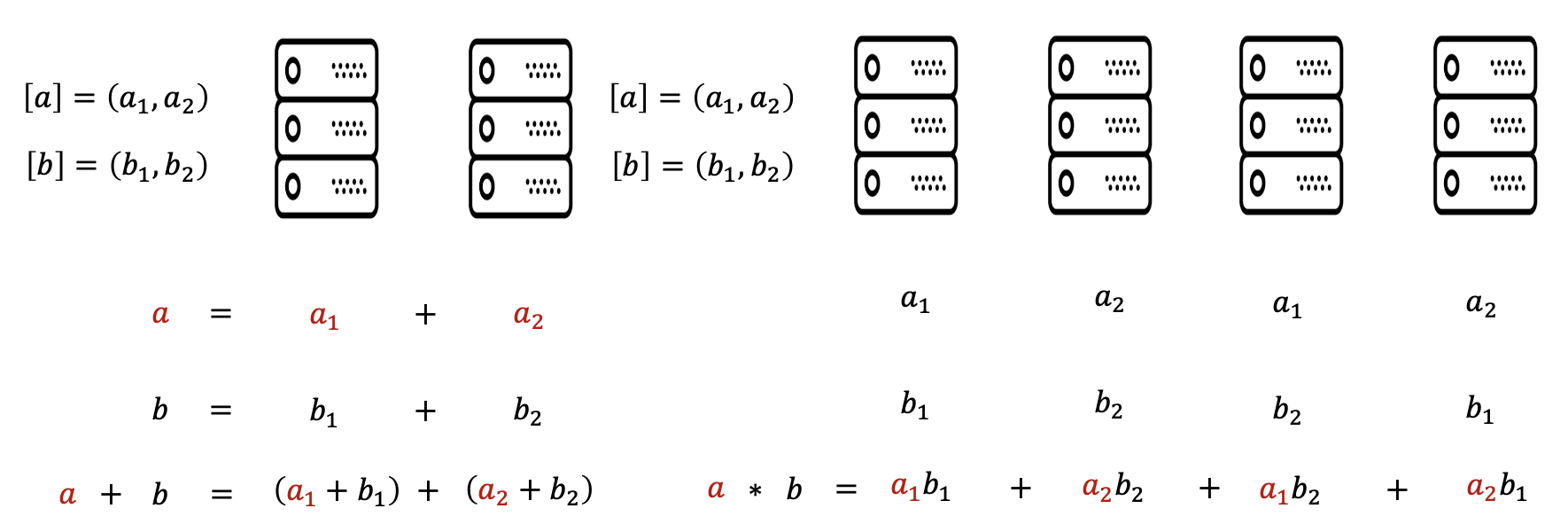}
    \caption{The Illustration of Secure Addition and Secure Multiplication}
    \label{fig:addmul}
\end{figure*}

As we know, the prediction model is constructed by the operation such as addition, multiplication, division, compare and so on. Those operation can be also implemented in the secure share protocol to protect the privacy \cite{li2019privpy}. The example of addition and multiplication is shown in Figure~\ref{fig:addmul}. \par

\begin{figure*}[htbp]
    \centering
    \includegraphics[width=0.8\textwidth]{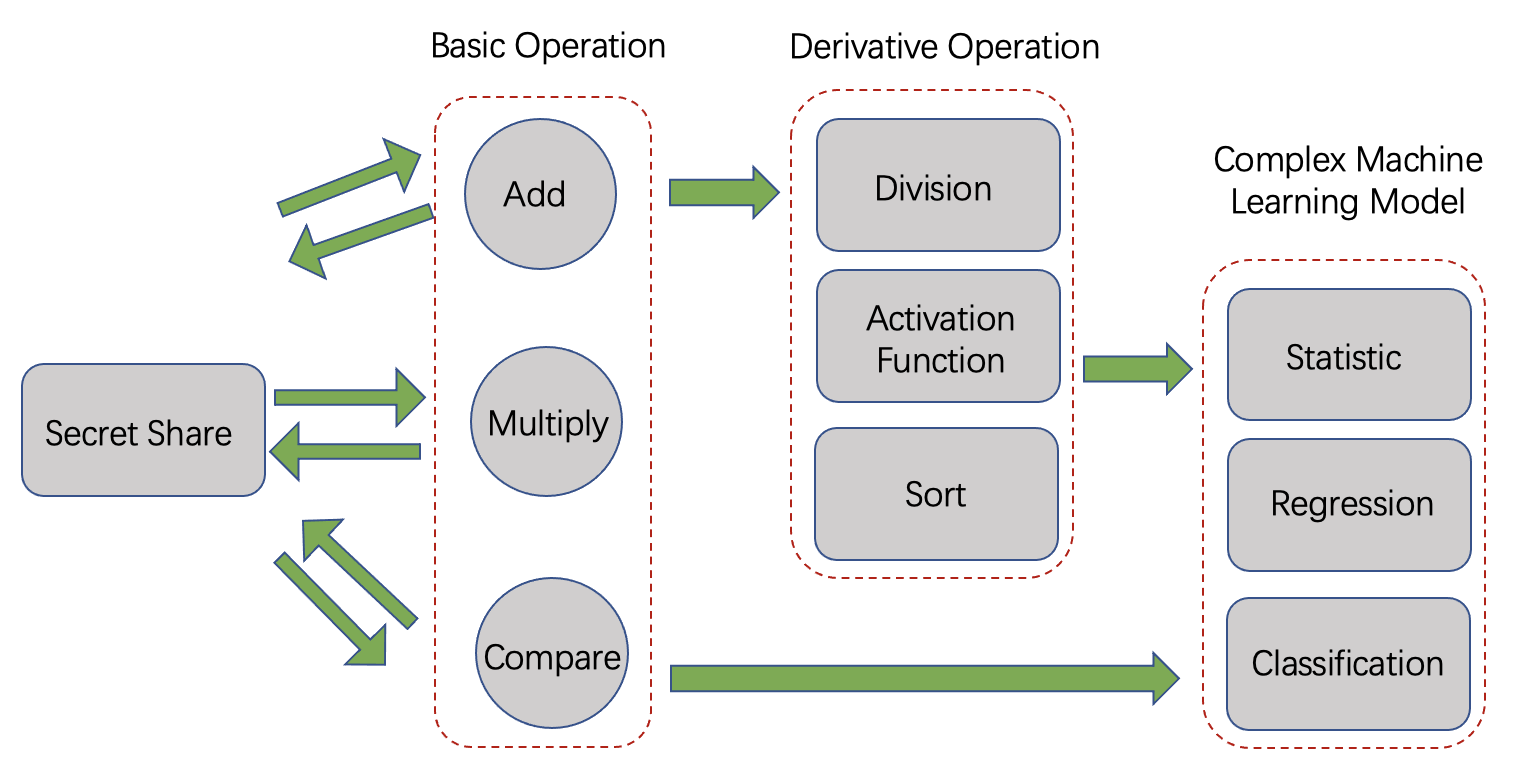}
    \caption{The Construction of Complex Machine Learning Model}
    \label{fig:construction}
\end{figure*}

For the secure addition, two part holds number $a$ and $b$ separately. According to secret share protocol, number $a$ is divided into two random number $a_{1}$ and $a_{2}$. Number $b$ is divided into two random number $b_{1}$ and $b_{2}$. Then $a_{1}$ and $b_{1}$ are sent to one computing server to get $a_{1}+b_{1}$. In the meanwhile, $a_{2}$ and $b_{2}$ are sent to another computing server to get $a_{2}+b_{2}$. By add the number of $a_{1}+b_{1}$ and $a_{2}+b_{2}$, the results of $a+b$ is worked out. If the two computing server will not collude, the privacy of $a$ and $b$ is also guaranteed. The computing process is similar for the secure multiplication. $a_{1}\times b_{1}$, $a_{2}\times b_{2}$, $a_{1}\times b_{2}$ and $a_{2}\times b_{1}$ are calculated separately on the computing server. Then the results of $a\times b$ can be worked out by adding those four components. If those four computing server not collude, the privacy can also be guaranteed. It The basic computation operation such as add, multiply and compare based on secret share is described in the Appendix~\ref{app:secmpc}. By utilizing the basic operation, we can construct the derivative operation such as division, activation function and sort. By leveraging the basic operation and derivative operation, we can build the complex machine learning method to approximate the wind power prediction function shown in equation (2). The process is shown in Figure~\ref{fig:construction}. 

\section{Privacy Preserving Ultra-short-term Wind Power Prediction} \label{sec:windpow}
\subsection{The Overview of Privacy Preserving Ultra-short-term Wind Power Prediction}
\begin{figure*}[htbp]
    \centering
    \includegraphics[width=0.9\textwidth]{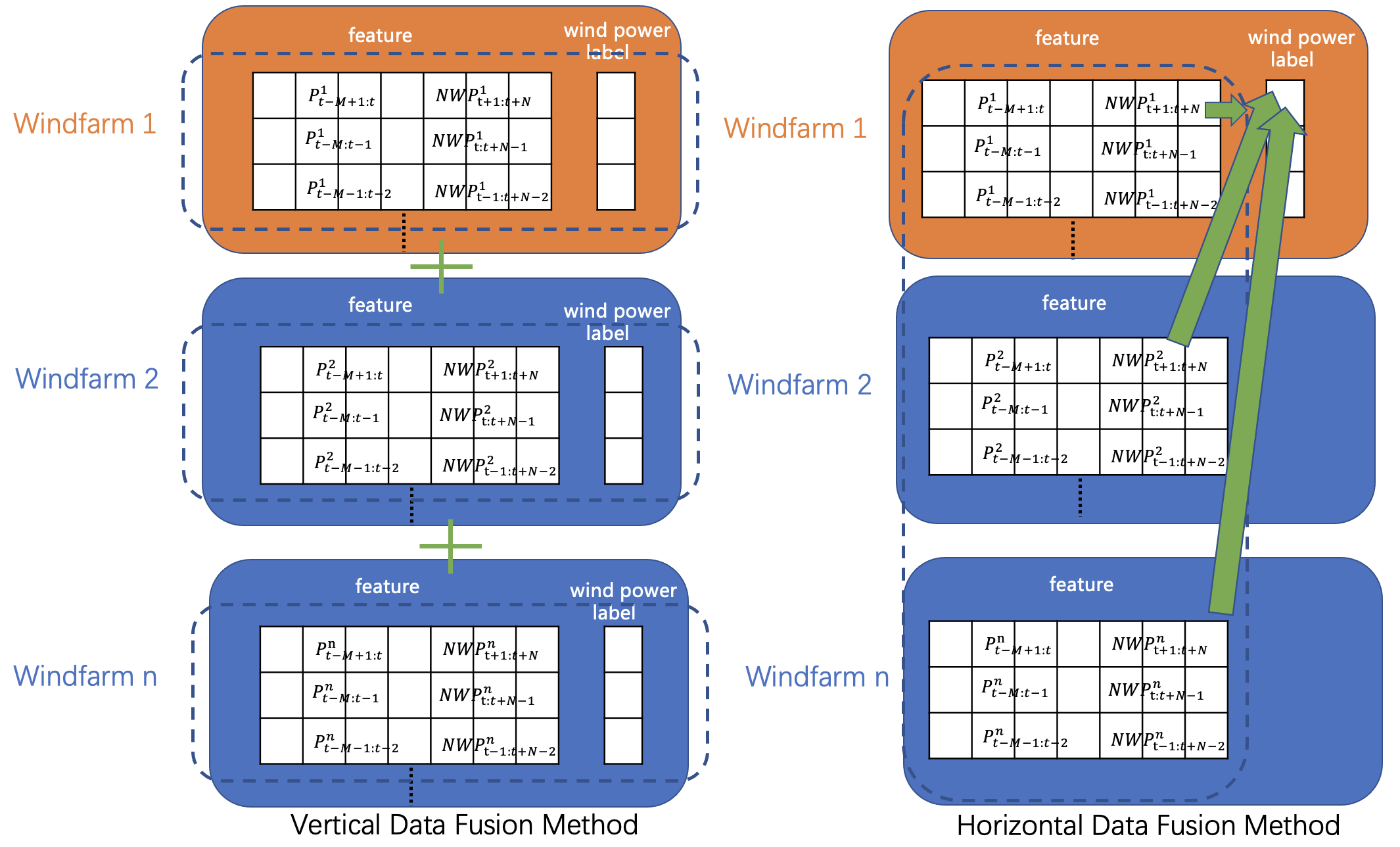}
    \caption{Horizontal and Vertical Data Fusion Method}
    \label{fig:horiver}
\end{figure*}

The basic idea of privacy preserving ultra-short-term wind power prediction is using the data of other wind farms to enhance the accuracy of prediction model. Indeed, not only the historical wind power but also the NWP data can be utilized in the privacy preserving training process. There are two ways of utilizing and fusing the data as shown in Figure~\ref{fig:horiver}. The first one is using the historical wind power, NWP and the labels of all the wind farms to train a model which is similar to the transfer learning \cite{zhuang2020comprehensive}. It is a horizontal data fusion method. The problem for this data fusion method is that the spatial and temporal relationship is not well considered. The second one incorporates the historical wind power and NWP of other wind farms at the same time to predict the label which resembles a centralized prediction method. It is a vertical data fusion method. In the wind power prediction tasks, the spatial and temporal correlation is included in the wind power and NWP at the same time, so the vertical data fusion method is more suitable for this case. \par

The participants are divided into \textbf{active parties} and \textbf{passive parties}. In wind power prediction, the active party is the wind farm who would like to predict the wind power using the data from other wind farms. They have both feature data and label data. The passive party is the wind farm who only has the feature data. It lends data to the active party. However, when the active party initiated a request for a prediction, the active party need to select the appropriate wind farms to act as the passive party. If too many wind farms take part in the training process, the massive communication will decrease the training efficiency, and the discrepancy of the sample distribution will also effect the prediction accuracy. If there is not enough wind farms in the training process, the spatial and temporal correlation pattern cannot be fully and accurately explored.\par

\begin{figure*}[htbp]
    \centering
    \includegraphics[width=1.0\textwidth]{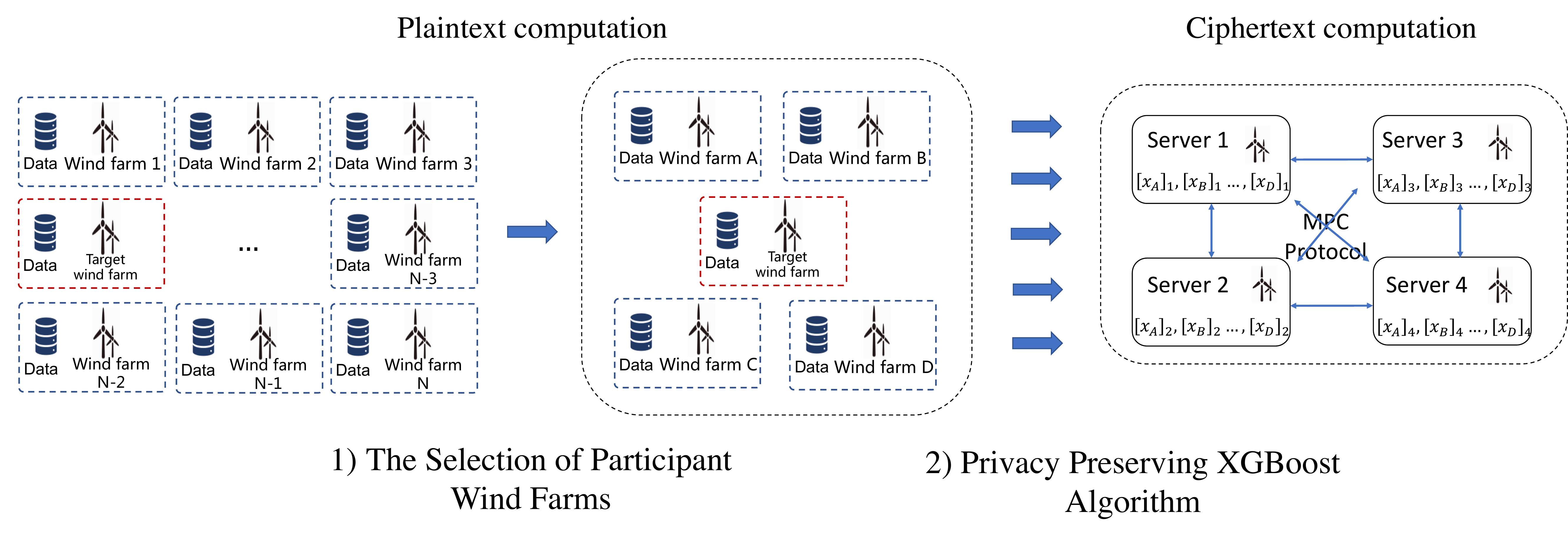}
    \caption{The Privacy Preserving Ultra-short-term Wind Power Prediction Process}
    \label{fig:predproc}
\end{figure*}

Therefore, in our pwXGBoost model, we divide the privacy preserving ultra-short-term wind power prediction process into two parts as shown in Figure~\ref{fig:predproc}. The first part is the selection of the participant wind farms (Section~\ref{sec:selection}) and the second part is the privacy preserving XGBoost algorithm for ultra-short-term wind power prediction (Section~\ref{sec:pwXGB}). 

\subsection{The Selection of Participant Wind Farms}\label{sec:selection}
In the privacy preserving ultra-short-term wind power prediction, it is necessary to select the wind farms which would like to engage in the data sharing. The reasonable selection of participant can increase the prediction accuracy.\par
Since the nearby wind farms have greater influence on the wind farm in the active party, we use the Gaussian distance function and adjacent matrix to determine which wind farm can be used in the prediction process. To make full use of the spatial and temporal correlation, the wind farms with similar distribution can be selected into a group and we use the \textit{Maximum Mean Discrepancy} (MMD) to calculate the distribution distance. MMD is widely used in the transfer learning which can measure the distance of two distribution. The multi kernel variant of MMD is used here which maps the data distributions into a \textit{Reproducing Kernel Hibert Space} (RKHS). For two distribution $d_1$ and $d_2$, the square of MMD between the two distributions can be calculated as \par

\begin{equation}
\begin{aligned}
\mathrm{MMD}^{2}(d_{1},d_{2}) = \left \|E\left[\phi(d_{1})\right]-E\left[\phi(d_{2})\right]\right \|
\end{aligned}
\end{equation}

Where $\phi(\cdot)$ is the mapping to RKHS. In practice, the mapping is unknown and we can use the kernel trick to replace the inner product. The result is 

\begin{equation}
\begin{aligned}
\mathrm{MMD}^{2}(d_{1},d_{2}) = E\left[K(d_{1},d_{1})\right] +E\left[K(d_{2},d_{2})\right] - 2E\left[K(d_{1},d_{2})\right]
\end{aligned}
\end{equation}

Where $K(d_{1},d_{2}) = \left[\phi(x),\phi(y)\right]$ is the desired kernel function. Because the wind farms are located in a region and the spatial and temporal correlation is closely related to the distribution distance, the adjacent matrix is defined as follows: \par

\begin{equation}\label{eq:11}
A_{i,j} = 
\begin{cases}
\exp\left(-\frac{\mathrm{MMD}^{2}(i,j)}{\mathrm{\sigma }^2}\right), &  \textrm{if $\mathrm{RMMD}^{2}(i,j)\leq \beta * \mathrm{mean}$} \\
0, &  \textrm{otherwise} \\
\end{cases}
\end{equation}

Where $\mathrm{MMD}^{2}(i,j)$ is the maximum mean discrepancy distance between wind farm $i$ and wind farm $j$, $\sigma$ is the standard deviation of the distance between $n$ wind farms and $\beta$ is the threshold. 
In our case, we set the half of the mean distance as the threshold. For wind farm $i$, only the wind farms whose $A_{i,j} \neq 0$ are chosen as the participants. \par

\subsection{The Privacy Preserving XGBoost Algorithm for Ultra-short-term Wind Power Prediction}\label{sec:pwXGB}
XGBoost is a method which is widely used in the wind power prediction\cite{zheng2019xgboost,li2020short}. There have been a number of widely adopted XGBoost programming packages, such as XGBoost \cite{chen2016xgboost}, LightGBM \cite{ke2017lightgbm} and CatBoost \cite{prokhorenkova2018catboost}. However, they are restricted to the centralized setting and are not applicable when data privacy is considered. Although there are some open-source privacy-preserving machine learning framework such as Pysytft \cite{ziller2021pysyft} and FedML \cite{he2020fedml}, they do not support XGBoost. Even though the FATE platform \cite{liu2021fate} can provide the XGBoost function, long encryption keys are required to ensure security. However, long keys can significantly slow down the computation, and the potential of privacy invading remains a problem. Therefore, it is necessary to build the XGBoost model based on secure multi-party computation which can fully preserve data privacy and is also effective in practice. \par

For the privacy preserving XGBoost algorithm for ultra-short-term wind power prediction, it can be divided into a training stage and a prediction stage. In the training stage, the historical wind power and NWP data of passive parties and the active party at time $t$ is fused in a vertical way as the feature, and the wind power to be predicted is the label. The privacy computation method is used because it can fuse the data from different sources without knowing their values. The prediction algorithm can be trained without compromising privacy. For the prediction stage, the encrypted historical wind power and NWP data are transmitted to the trained model and output the predicted wind power. This part is also shown in Figure~\ref{fig:predproc}.\par
According to the secure multi party computation theory, some wind farms are selected out as the computing server node randomly. The secret share of the data of each wind farm rather than the original data are sent to the computing server node. In this way, if the computing server node not collude, the security is guaranteed. We will discuss the details of the privacy preserving XGBost model based on secret sharing in Section ~\ref{sec:xgboost}.

\section{Privacy Preserving XGBoost Algorithm Based on Secret Sharing} \label{sec:xgboost}
In this section, we present the design and details of privacy preserving XGBoost Algorithm based on 2-out-of-4 secret sharing protocol. 

\subsection{Overall Description of Privacy Preserving XGBoost Model}

\begin{figure*}[htbp]
    \centering
    \includegraphics[width=0.8\textwidth]{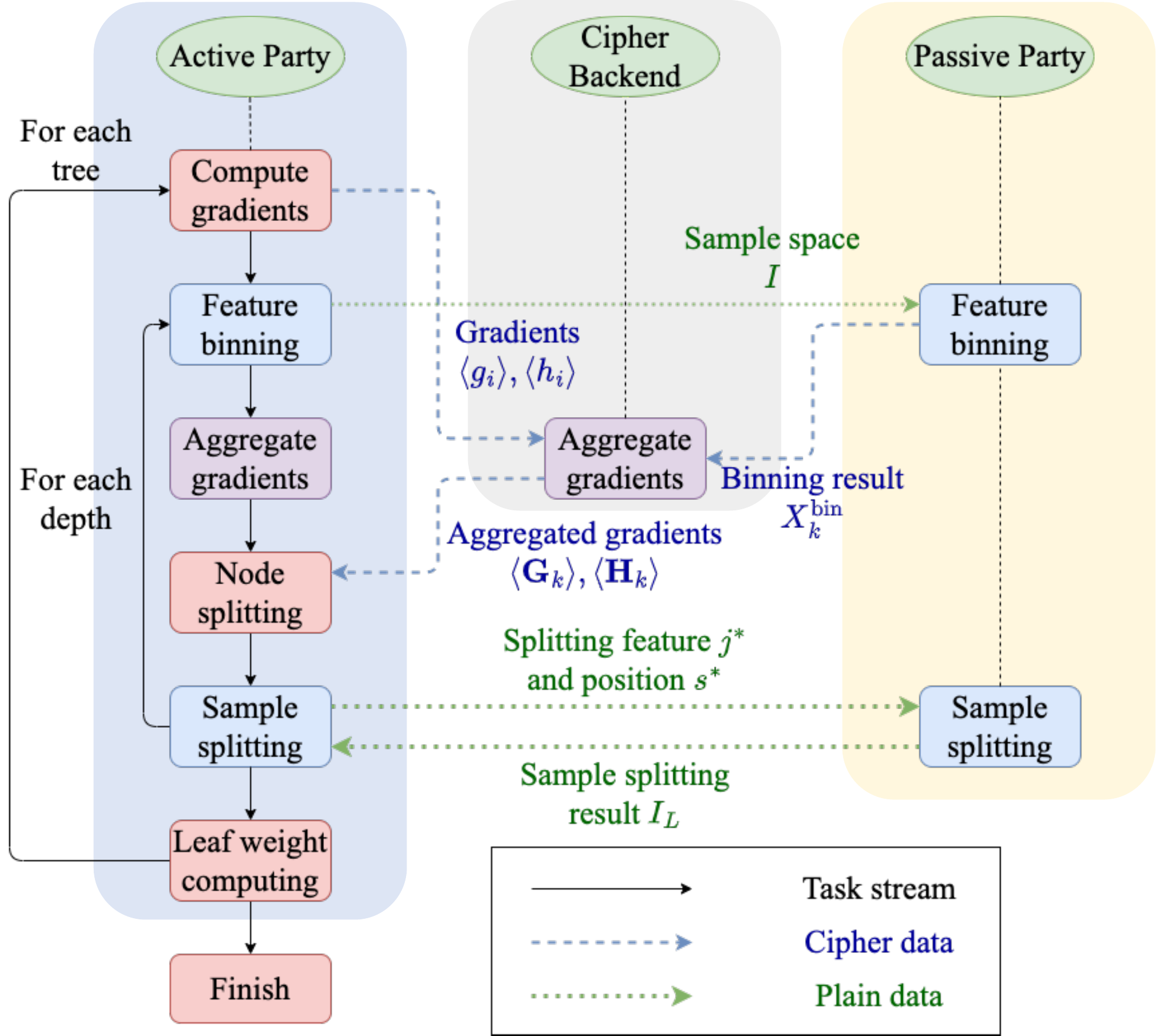}
    \caption{The Training Process of Privacy Preserving XGBoost in Secret Sharing}
    \label{fig:train}
\end{figure*}

The whole training process of the secure XGBoost model is shown in Figure~\ref{fig:train}. For the vertical privacy preserving XGBoost model, there are $m$ participants take part in the computation. Here we note the active party to be party $1$, and the passive parties to be party $2,\cdots,m$.

\subsubsection{Secure Model Training Process}

\begin{algorithm}[htbp]
	\caption{The main process of training a federated XGBoost model}
	\label{alg:alg1}
	\begin{algorithmic}[1]
		\REQUIRE On party $k$ ($1\le k\le m$): $\{\mathbf{X}_{k}\}_{M\times N_k}$, the feature dataset of the $k$-th party
            \REQUIRE On active party: $\{y\}_{M}$, the truth label dataset
            \REQUIRE $T$, number of trees; $D$, maximum depth of tree splitting; $B$, number of bins; $\gamma$, minimum loss reduction for a split; $\lambda$, L2 regularization term
		\ENSURE  The $t$-th decision tree for $1\le t \le T$. The main tree structure is stored on the active party, while the binning boundaries are stored distributed across all parties.

            \FOR{$1\le t\le T$}

            \STATE{$\hat{y}_i\leftarrow\sum_{i=1}^{t-1}f_i(x_i)$}

            \STATE{$g_i \leftarrow -y_{i}+\frac{1}{1+e^{-\hat{y}_{i}}}$, $h_i \leftarrow \frac{e^{-\hat{y}_{i}}}{(1+e^{-\hat{y}_{i}})^2}$}\;
            
            \STATE{Active party: Encrypt all gradients $g_i$ and $h_i$. Send all $[ g_i]$ and $[ h_i]$ to the ciphertext side.}

            \STATE{Initialize a new tree, add the root node $u_0$ to it}
            \STATE{For root node $u_0$, set the sample space $I_{u_0}\leftarrow\{1,2,\cdots,M\}$}
		
		\FOR{$1\le d \le D$} 
            \FOR{each tree node on depth $d-1$ (parallelizable on all nodes)}
			\STATE{Active party: For this tree node, send its sample space $I$ to all passive parties}\;
                \FOR{$1\le k \le M$ (parallelizable on all $k$)}
                \STATE{Party $k$: Do binning on $I$ and $\mathbf{X}_k$, getting the binning result $\mathbf{X}^{\mathrm {bin}}_{k}$ and binning boundaries $b_{k,j}$ for $1\le j\le N_k$ (Algorithm \ref{alg:alg2}). Encrypt $\mathbf{X}^{\mathrm {bin}}_{k}$ and send $[\mathbf{X}^{\mathrm {bin}}_{k}]$ to the ciphertext side. Save the binning boundaries $b_{k,j}$ to local storage for $1\le j\le N_k$.}\;
                \STATE{Ciphertext Side: According to $[ g_i]$, $[ h_i]$ and $[\mathbf{X}^{\mathrm {bin}}_{k}]$, compute the aggregated gradients $[\mathbf{G}_k]$ and $[\mathbf{H}_k]$ (Algorithm \ref{alg:alg3}). Send the aggregated gradients $[\mathbf{G}_k]$ and $[\mathbf{H}_k]$ to the active party. For the active party ($k=1$), this process can be computed in pure plain text on the active party itself instead.}\;
                \ENDFOR

                \STATE{Active party: Decrypt the aggregated gradients $[\mathbf{G}_k]$ and $[\mathbf{H}_k]$ for all $k$, get $\mathbf{G}_k$ and $\mathbf{H}_k$. Compute the best split $(k^*, j^*, s^*)$ of this tree node (Algorithm \ref{alg:alg4}), where $k^*$ is the split party, $j^*$ is the split feature ID on party $k^*$, and $s^*$ is the position of split point on feature $j^*$. Send the tuple $(j^*, s^*$ to party $k^*$. (It is possible that the split gain is not greater than $0$. In that case, continue to the next node without splitting the tree.)}
			
			\STATE{Party $k^*$: According to $(j^*$ and $s^*)$, for sample vector $I$, determine the left sample space $I_{L}$ (Algorithm \ref{alg:alg5}). Send $I_{L}$ to the active party.}
			
			\STATE{Active Party: Compute $I_R\leftarrow I-I_L$. Split the current tree node into two child nodes to join the node queue, assign $I_L$ and $I_R$ to them respectively.}
			
		\ENDFOR
            \ENDFOR
            \FOR{each leaf node $u$ in the tree}
            \STATE{Compute the weight $w_u \leftarrow -\frac{\sum_{i\in I_{u}}g_{i}}{\sum_{i\in I_{u}}h_{i}+\lambda}$.}
            \ENDFOR
            \STATE{Add the new generated decision tree to the model.}
            \ENDFOR
            \RETURN the generated model with all trees
	\end{algorithmic}
\end{algorithm}

The algorithm of the secure XGBoost model is shown in Algorithm~\ref{alg:alg1}. The model is trained among the active party, passive party and the ciphertext backend. 
The input datasets of the algorithm including the feature set $\{\mathbf{X}_{k}\}_{M\times N_k}$ ($1\le k\le m$) provided by all the $m$ parties and the label set $\{y\}_{M}$ owned by the active party. 
During the training process, we train the $T$ decision trees sequentially. 

For the $t$-th tree, at the beginning, we compute the prediction result $\hat{y}_i$ ($1\le i\le M$) of the first $t-1$ trees. Then we compute the first and second order gradients, which are $g_i$ and $h_i$ ($1\le i\le M$) respectively. We encrypt them and send $[ g_i]$ and $[ h_i]$ ($1\le i\le M$) to the cipher end. For each tree, we are specified the value $D$ which is the max depth of tree. We build the root node $u_0$ at first, and set the sample space $I_{u_0}$ to be the set of all samples. Then we split from the root node depth by depth. For each node at a specified depth, the active party send its sample space $I$ to all passive parties. Then for each party $k$, do binning on the feature dataset $\mathbf{X}_k$ and sample space $I$. Encrypt the binning result as $\left[ \mathbf{X}_k^{\mathrm{bin}} \right]$ and send it to the cipher and, and save the binning boundaries to local storage. On the cipher end, for each party, aggregate the encrypted gradients based on the encrypted binning results, and send the encrypted $[\mathbf{G}_k]$ and $[\mathbf{H}_k]$ to the active party. On the active party, decrypt the aggregated gradients and compute the information gain for each party and each feature, and then decide the optimal split $(k^*, j^*, s^*)$. Send the optimal split to the corresponding party, to compute the sample space of the child nodes. Repeat this process until the maximum depth is reached, and then compute the weight of each leaf node. After that, add the new generated tree to the model.

\subsubsection{Secure Gradient Computation}

\begin{algorithm}[htbp]
	\caption{Local binning algorithm on Party $k$} 
	\label{alg:alg2}
	\begin{algorithmic}[1]
		\REQUIRE $I$, sample space of the current tree node (a list composed of sample indexes)
		\REQUIRE $\{\mathbf{X}_k\}_{M\times N_k}$, feature dataset of party $k$.
            \REQUIRE $B$, number of bins.
		\ENSURE $\mathbf{X}^{\mathrm {bin}}_{k}$, binned result of party $k$; $b_{k,j}$ ($1\le j\le N_k$), binning boundary vectors

        \STATE $\mathbf{X}^{\mathrm{bin}}_k\leftarrow\{-1\}_{M\times N_k}$
        \FOR {$1\le j \le N_k$}
        \STATE Process a quantile binning on the $j$-th column of $\mathbf{X}_k$ based on the sample space $I$, where the number of bins is $B$. Assume the vector $\{x^{\mathrm{bin}}_j\}_M$ to be the binning result, and $b_{k,j}$ is the binning boundary vector. $x^{\mathrm{bin}}_j$ is a vector of length $m$, where the values are between $0$ and $M-1$. $b_{k,j}$ is a vector of length $B-1$.
        \FOR {$1\le i\le M$}
            \IF{$i\in I$}
            \STATE $\{\mathbf{X}^{\mathrm{bin}}_k\}_{i,j}\leftarrow\{x^{\mathrm{bin}}_{j}\}_i$
            \ENDIF
        \ENDFOR
        \ENDFOR
        \RETURN{$\mathbf{X}^{\mathrm {bin}}_{k}$, $b_{k,j}$ ($1\le j\le N_k$)}
	\end{algorithmic}

\end{algorithm}

The local binning algorithm is shown in Algorithm~\ref{alg:alg2}. On each active or passive party, after getting the feature dataset $\{\mathbf{X}_{k}\}_{M\times N_k}$, process a quantile binning algorithm on each column based on the sample space $I$. The binning result $\mathbf{X}^{\mathrm {bin}}_{k}$ is encrypted and send to the cipher end, while the binning boundaries are saved to local storage.

\begin{algorithm}[htbp]
	\caption{Cipher gradient aggregation of party $k$ on the ciphertext side (for the active party, this algorithm can be run in pure plain text on the active party)} 
	\label{alg:alg3}
	\begin{algorithmic}[1]
            \REQUIRE {$\left [ g_i \right]$ and $\left [ h_i\right]$ ($1\le i\le M$), encrypted gradient vectors}
		\REQUIRE $[\mathbf{X}^{\mathrm {bin}}_{k}]_{M\times N_k}$, encrypted binned result of party $k$
		\ENSURE {$\left [ \mathbf{G}_{k} \right], \left [ \mathbf{H}_{k} \right]$, encrypted aggregated gradients for party $k$}

        \STATE $\left [ \mathbf{G}_{k} \right]\leftarrow[0]_{N_k\times B}, \left [ \mathbf{H}_{k} \right]\leftarrow[0]_{N_k\times B}$
        \FOR{$1\le j \le N_k$}
        \FOR{$1\le b \le B$}
        \FOR{$1\le i \le M$}
        \STATE $[\mathrm{flag}]\leftarrow[\mathbf{X}^{\mathrm {bin}}_{k}]_{i,j}=b$
        \STATE $[ \mathbf{G}_{k} ]_{j,b} \leftarrow [ \mathbf{G}_{k} ]_{j,b} + [\mathrm{flag}] * \left [ g_i \right] $
        \STATE $[ \mathbf{H}_{k} ]_{j,b} \leftarrow [ \mathbf{H}_{k} ]_{j,b} + [\mathrm{flag}] * \left [ h_i \right] $
        \ENDFOR
        \ENDFOR
        \ENDFOR

        \RETURN{$\left [ \mathbf{G}_{k} \right], \left [ \mathbf{H}_{k} \right]$}

	\end{algorithmic}

\end{algorithm}

The secure gradient aggregation algorithm is shown in Algorithm~\ref{alg:alg3}. For each feature $j$ and each bin $b$, find out the set of samples which appear in that bin. To preserve security, the process of dividing samples into bins is still in an encrypted computation, and the result is an encrypted $0$-$1$ vector $[\mathrm{flag}]$. After that, aggregate the gradients with the weight  $[\mathrm{flag}]$. Send the aggregated gradients $[\mathbf{G}_k]$ and $[\mathbf{H}_k]$ to the active party.

\subsubsection{Secure Split}

\begin{algorithm}[htbp]
	\caption{Compute the optimal split point on the active party} 
	\label{alg:alg4}
	\begin{algorithmic}[1]
		\REQUIRE $I$, sample space of the current tree node (a list composed of sample indexes)
		\REQUIRE {$\{\mathbf{G}_{k}\}_{N_k\times B}, \{\mathbf{H}_{k}\}_{N_k\times B}$, aggregated gradients from all parties $1\le k\le m$}
		\REQUIRE $\gamma$, minimum loss reduction for a split; $\lambda$, L2 regularization term

            \ENSURE $k^*$, party ID of the optimal split; $j^*$, feature ID of the optimal split on party $k^*$; $s^*$, position of the optimal split on feature $j^*$ of party $k^*$

        \STATE $k^*\leftarrow \mathrm{null}$, $j^*\leftarrow \mathrm{null}$, $s^*\leftarrow \mathrm{null}$, $v^*\leftarrow 0$
        \FOR{$1\le k\le m$}
        \FOR{$1\le j\le N_k$}
        \STATE $G\leftarrow\sum_{b=1}^B\{\mathbf{G}_k\}_{j,b}, H\leftarrow\sum_{b=1}^B\{\mathbf{H}_k\}_{j,b}$
        \STATE{$G_L\leftarrow0, G_R\leftarrow G$, $H_L\leftarrow0, H_R\leftarrow H$}
        \FOR{$1\le s\le B-1$}
        \STATE $G_L\leftarrow G_L+\{\mathbf{G}_k\}_{j,s}, H_L\leftarrow H_L+\{\mathbf{H}_k\}_{j,s}$
        \STATE $G_R\leftarrow G-G_L, H_R\leftarrow H-H_L$
        \STATE $ v \leftarrow \frac12\left(\frac{G_{L}^2}{H_{L}+\lambda} + \frac{G_{R}^2}{H_{R}+\lambda} - \frac{G^2}{H +\lambda}\right)-\gamma$
        \IF{$v>v^*$}
        \STATE $k^*\leftarrow k$, $j^*\leftarrow j$, $s^*\leftarrow s$, $v^* \leftarrow v$
        \ENDIF
        \ENDFOR
        \ENDFOR
        \ENDFOR
        \RETURN $k^*,j^*,s^*$
	\end{algorithmic}
\end{algorithm}

The tree node splitting algorithm is shown in Algorithm~\ref{alg:alg4}. On the active party, after collecting all encrypted aggregated gradients from the cipher end, decrypt them to get $\mathbf{G}_k$ and $\mathbf{H}_k$ ($1\le k\le m$). Then for each party, each feature and each split position, compute the information gain based on aggregated gradients. Unless the information gains are non-positive for all splits, find the optimal split $(k^*,j^*,s^*)$ with the best information gain $v^*$, and split the tree node to two child nodes. 

\begin{algorithm}[htbp]
	\caption{Sample splitting on party $k_{\mathrm{opt}}$} 
	\label{alg:alg5}
	\begin{algorithmic}[1]
		\REQUIRE $I$, sample space of the current tree node (a list composed of sample indexes)
		\REQUIRE $j^*$, feature ID of the optimal split on party $k^*$; $s^*$, position of the optimal split on feature $j^*$ of party $k^*$
		\REQUIRE $\{\mathbf{X}_{k^*}\}_{M\times N_{k^*}}$, feature dataset of party $k^*$.
            \REQUIRE $\{b_{{k^*},j^*}\}_{B-1}$, binning boundaries of the $j^*$-th feature.

            \ENSURE $I_L$, sample space of the left child node

        \STATE $I_L\leftarrow\{\}$
        \FOR{$i\in I$}
        \IF{$\{\mathbf{X}_{k^*}\}_{i, j^*}\le\{b_{{k^*},j^*}\}_{s^*}$}
        \STATE $I_L\leftarrow I_L\cup\{i\}$
        \ENDIF
        \ENDFOR
        \RETURN $I_L$
	\end{algorithmic}
\end{algorithm}

The sample splitting algorithm is shown in Algorithm~\ref{alg:alg5}. After party $k^*$ receives the optimal split $(j^*,s^*)$, it compares the corresponding feature values in the sample space with the binning boundary of the optimal split, and decide whether each sample should be allocated to the left or right child node. After getting the result, return the left sample space $I_L$ to the active party.

\section{Security Analysis} \label{sec:secu}

\paragraph{Security assumptions.  } 
For scalability and generality, we model all the participant $N$ wind farms as \textit{data providers} who agree to contribute their data for power prediction in a privacy-preserving way, and for any $t$-out-of-$n$ secret sharing protocols, we further select $n$ wind farms as the \textit{computation servers} to carry the ciphertext computation, where $n \leq N$ and $t < n$. The $t$-out-of-$n$ secret sharing protocol guarantees that any set of $t$ computation servers together can reconstruct the raw data while any set less then $t$ servers learns nothing.
The computation servers are connected through \textit{secure channels}, hold the secret shares of all the data providers data and execute the agreed secure protocols. Through this model, our method can be extended to any number wind farms with willingness to share their data and suitable for any $t$-out-of-$n$ secret sharing protocols. 

Similiar to other main-stream privacy-preserving applications~\cite{ben2016optimizing, patel2015efficient, fan2021ppca}, we define our security model as \textit{honest-majority} and \textit{semi-honest} model~\cite{evans2018pragmatic} for practical performance. The above model assumes that during the ciphertext computation, no more than a half computation servers are corrupted together ($t < \frac{n}{2}$) and the corrupted servers will follow the agreed protocol while try to learn as much information as possible about the others. 
Informally, a protocol is secure in the above model if the information that the corrupted servers gained is not distinguishable as there exists an ideal trusted third party.  

\paragraph{Security analysis.  } The computation of our method can be separated into two parts, plaintext computation and ciphertext computation. 
The security of ciphertext computation is guaranteed through the \textit{modular composition} theorem~\cite{canetti2000security}, which offers a general way for designing complex high-level secure protocols. Firstly, we design the high-level protocol by assuming that a series of simple sub-protocols exist. Then we design each sub-protocol meeting the security guarantee and then plug them as sub-routines in the high-level protocol. The modular composition theorem formally stated that, if the high-level protocol can be securely evaluated its function with ideal protocols, then the security and functionality maintained by replacing all the ideal sub-protocols with sub-routines~\cite{canetti2000security}. 

For all the participant wind farms, the data providers only do plaintext computation of their own data, thus introducing no privacy risk. The computation servers only see the secret shares of the raw data and carry all the ciphertext computation. In our method, all the cross party computation are designed to be evaluated in the ciphertext, and all the ciphertext computation logic is modular compositions of secure addition, subtraction, multiplication, comparison, reciprocal as well as exponential, which are well-studied and commonly provided by most semi-honest MPC platforms like \cite{li2019privpy,keller2020mp}. Thus, the security of each wind farm's data is preserved.

Note that he participants selection stage only utilizes 2 week history data of each wind farm to select the most similar participants in plaintext as the correlation among wind farms in a specific season is relatively stable and the MMD calculation is computationally intensive. In practice, this method performs effectively.






\section{Case Studies} \label{sec:case}
The proposed privacy-preserving prediction method is tested on the field measurement data of a wind farm cluster in Inner Mongolia, China. The privacy preserving machine learning algorithms are implemented on the PrivPy, a general-purpose MPC platform~\cite{li2019privpy} which offers a series of secure operations based on 2-out-of-4 secret sharing protocol.
All the evaluations are performed on a Kubernetes (k8s) cluster~\cite{burns2022kubernetes} that is deployed on two 64-core AMD EPYC CPUs with 256GB RAM. Each wind farm and computing server is deployed as a separate k8s container, and the round-trip time between each pair is approximately 0.1ms.

\subsection{Data Set and Test Description}
\begin{figure}[htbp]
    \centering
    \includegraphics[width=0.7\textwidth]{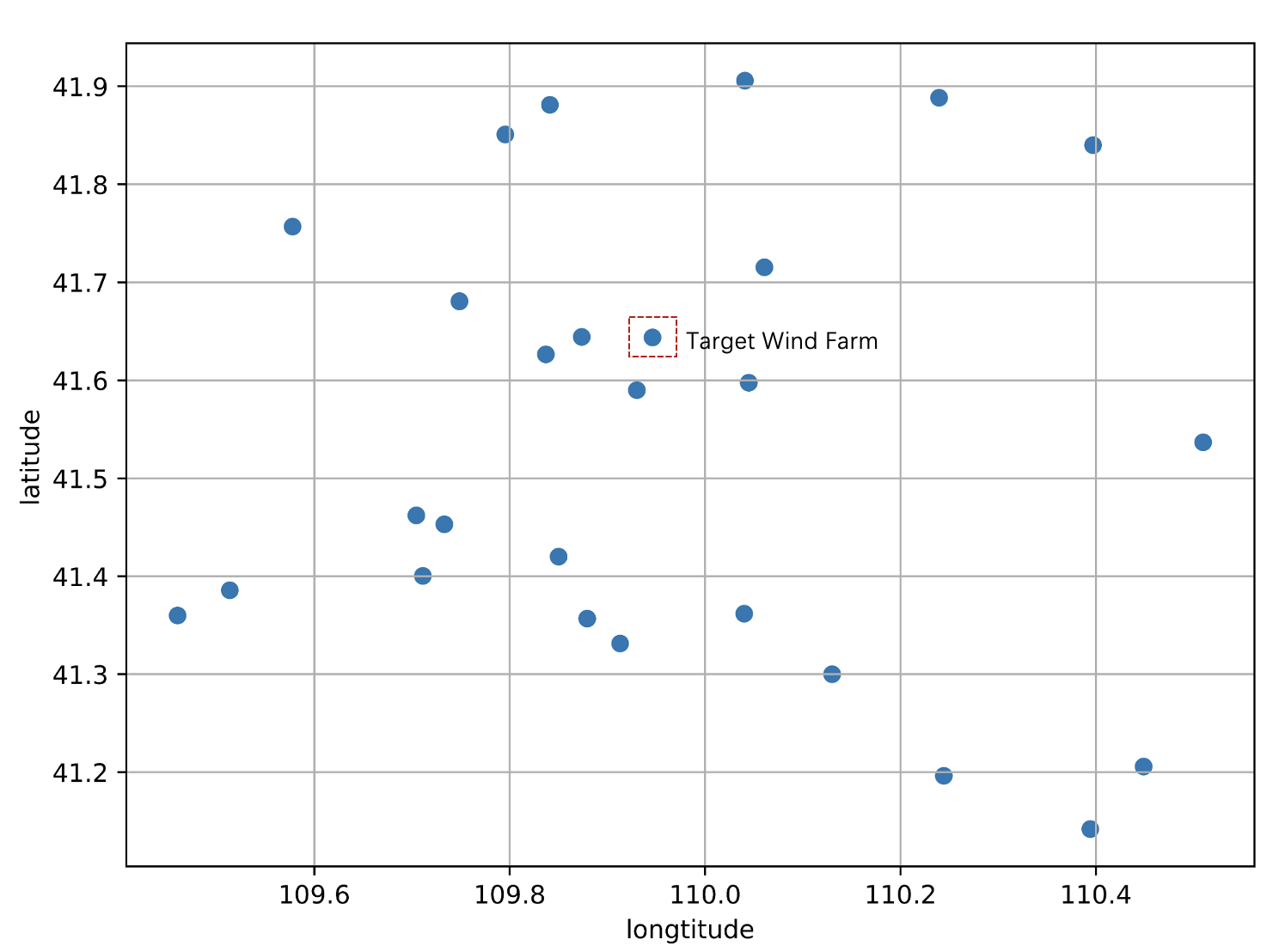}
    \caption{The Location of Wind Farms}
    \label{fig:loc}
\end{figure}

The wind power and NWP data from Jan. 1st 2021 to July 23th 2021 are recorded with a time step of 15 minutes. The locations of 27 wind farms are shown in Figure~\ref{fig:loc}. We will use a wind farm located in the center of the chosen cluster as the target wind farm for example to illustrate the effectiveness of privacy preserving collaborative prediction model, pwXGBoost. \par

To illustrate the effectiveness of the pwXGBoost, the root mean square error (RMSE)

\begin{equation}
\begin{aligned}
\mathrm{RMSE} = \sqrt{\frac{1}{k}\sum_{i=1}^{k}(x_{ti}-\hat{x}_{ti})^2}
\end{aligned}
\end{equation}

 and the mean absolute error (MAE)
 
\begin{equation}
\begin{aligned}
\mathrm{MAE} = \frac{1}{k}\sum_{i=1}^{k}\left|x_{ti}-\hat{x}_{ti}\right|
\end{aligned}
\end{equation}

 are used as the indexes to assess the prediction accuracy on the dataset, where $k$ is the number of the samples in the dataset. $x_{ti}$ is the $i$ th wind power sample at time $t$ and $\hat{x}_{ti}$ is the predicted $i$ th wind power sample at time $t$.

\subsection{The Prediction Results of the Privacy Preserving Collaborative Prediction Model}
The privacy preserving XGBoost model is the method proposed in this paper for the ultra-short-term wind power prediction. The key advantage of this method is that it can not only utilize the historical wind power and NWP data from other wind farms without breaching the privacy, but also can extract the nonlinear spatial and temporal relationship compared to the linear method before. To illustrate the effectiveness of the method, we compare it with the methods which only utilize the local data and those use the linear model to extract the spatial and temporal correlation. 
\paragraph{Local\_XGBoost\_wo\_nwp} This model only uses the local historical wind power data to train the XGBoost model. The max depth of the tree in the model is 3, learning rate is 0.3 and the tree number which is the estimator of the XGBoost is 80.
\paragraph{Local\_XGBoost} This model only uses the local historical wind power data and NWP data to train the XGBoost model. The max depth of the tree in the model is 3, learning rate is 0.3 and the tree number which is the estimator of the XGBoost is 80.
\paragraph{Lasso\_wo\_nwp} \cite{goncalves2021privacy} This model can use the wind power from nearby wind farms but is the linear method. The alpha parameter is 0.00005.
\paragraph{Lasso} \cite{goncalves2021privacy} This model can use the wind power and NWP data from nearby wind farms but is the linear method. The alpha parameter is 0.00005.
\paragraph{pwXGBoost\_wo\_nwp\_mmd} This model can use the historical wind power from nearby wind farms. The nonlinear spatial and temporal relationship can be extracted. The max depth of the tree in the model is 3, learning rate is 0.3 and the tree number which is the estimator of the XGBoost is 80.
\paragraph{pwXGBoost\_wo\_mmd} This model can use the historical wind power and NWP data from nearby wind farms. The nonlinear spatial and temporal relationship can be extracted. The max depth of the tree in the model is 3, learning rate is 0.3 and the tree number which is the estimator of the XGBoost is 80. The correlated wind farms are selected based on the distance.
\paragraph{pwXGBoost} This model can use the historical wind power and NWP data from nearby wind farms. The nonlinear spatial and temporal relationship can be extracted. The max depth of the tree in the model is 3, learning rate is 0.3 and the tree number which is the estimator of the XGBoost is 80. The correlated wind farms are selected based on the MMD.
The Prediction results of different models are as in Table~\ref{tab:result}. \par

\begin{table*}[htbp]
\caption{The Prediction Results of Different Models (\%)}\label{tab:result}
\center
\begin{tabular}{|c|c|c|c|c|c|c|c|c|}\hline
\multicolumn{1}{|c|}{\multirow{2}{*}{Method}} & \multicolumn{2}{c|}{1h}& \multicolumn{2}{c|}{2h} & \multicolumn{2}{c|}{3h} & \multicolumn{2}{c|}{4h}\\ \cline{2-9}
 & RMSE   & MAE & RMSE  & MAE & RMSE  & MAE & RMSE  & MAE\\ \hline
Local\_XGBoost\_wo\_nwp & 4.035  & 2.678 & 4.876  & 3.403 & 5.426  & 3.938 & 5.554  & 3.920 \\ \hline
Local\_XGBoost & 4.033  & 2.673 & 4.843  & 3.424 & 5.301  & 3.758 & 5.872  & 4.354 \\ \hline
Lasso\_wo\_nwp & 3.823  & 2.462 & 4.629  & 3.289 & 5.293  & 3.729 & 5.552  & 4.224 \\ \hline
Lasso & 3.817  & 2.401 & 4.583  & 3.181 & 5.243  & 3.881 & 5.790  & 4.269 \\ \hline
pwXGBoost\_wo\_nwp\_mmd & 3.840  & 2.480 & 4.649 & 3.225  & 5.183 & 3.662  & 5.609  & 4.100\\ \hline
pwXGBoost\_wo\_mmd & 3.812  & 2.443 & 4.536  & 3.167 & 5.232  & 3.723 & 5.553 & 3.920 \\ \hline
pwXGBoost & \textbf{3.781}  & \textbf{2.346} & \textbf{4.409}  & \textbf{2.962} & \textbf{4.923}  & \textbf{3.453} & \textbf{5.325} & \textbf{3.761} \\ \hline
\end{tabular}

\end{table*}

\begin{figure}[htbp]
    \centering
    \includegraphics[width=0.9\textwidth]{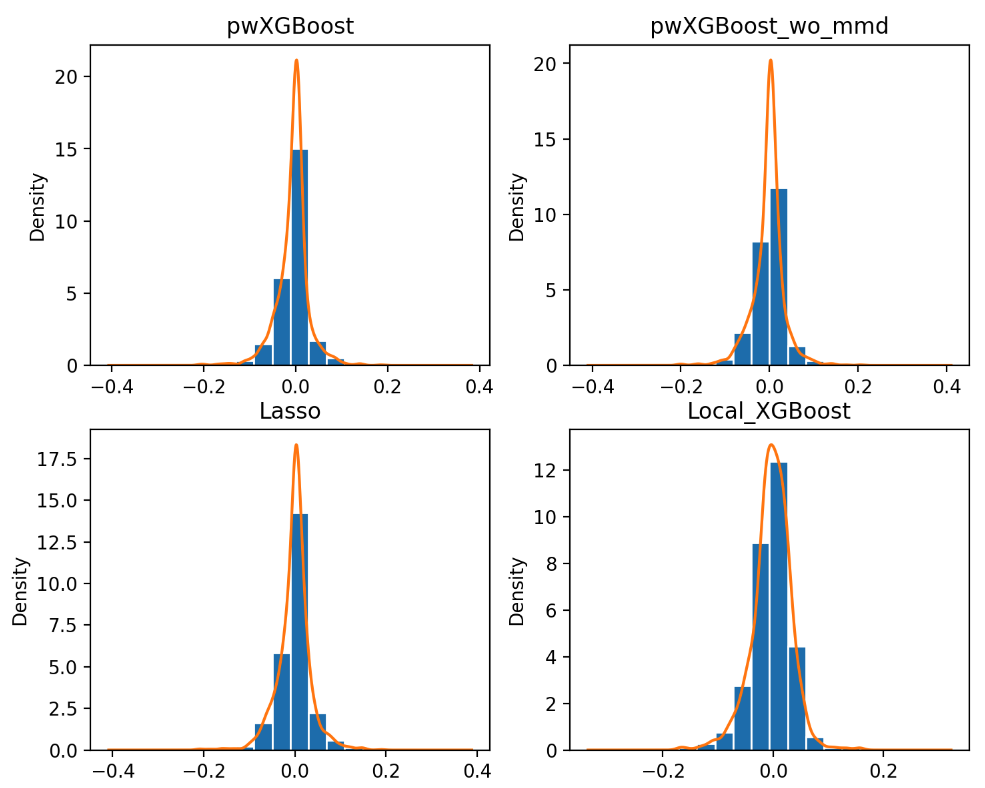}
    \caption{Probability distribution of forecasting errors}
    \label{fig:compres}
\end{figure}

According to the results in Table~\ref{tab:result}, the proposed pwXGBoost is better than the linear method especially when the prediction time is longer. Because when the prediction horizontal increase, the nonlinear of the wind power also increase and the superiority of pwXGBoost method is demonstrated. Besides, methods that take advantage of spatial and temporal correlation patterns always performs better than those which do not exploit the correlation. It is found that MMD is more effective when selecting the correlated wind farms than using the wind farm distance. We also visualize the error density of the prediction model for the $4$th hour in Figure~\ref{fig:compres}. \par

In Figure~\ref{fig:compres}, the prediction error is plotted in hist gram and fitted by kernel density estimation method. It is proved that the prediction error of the pwXGBoost is more centralized around zero which means it has better prediction performance. 

\subsection{The Analysis of the Privacy Preserving Prediction Results}
\begin{figure}[htbp]
    \centering
    \includegraphics[width=0.8\textwidth]{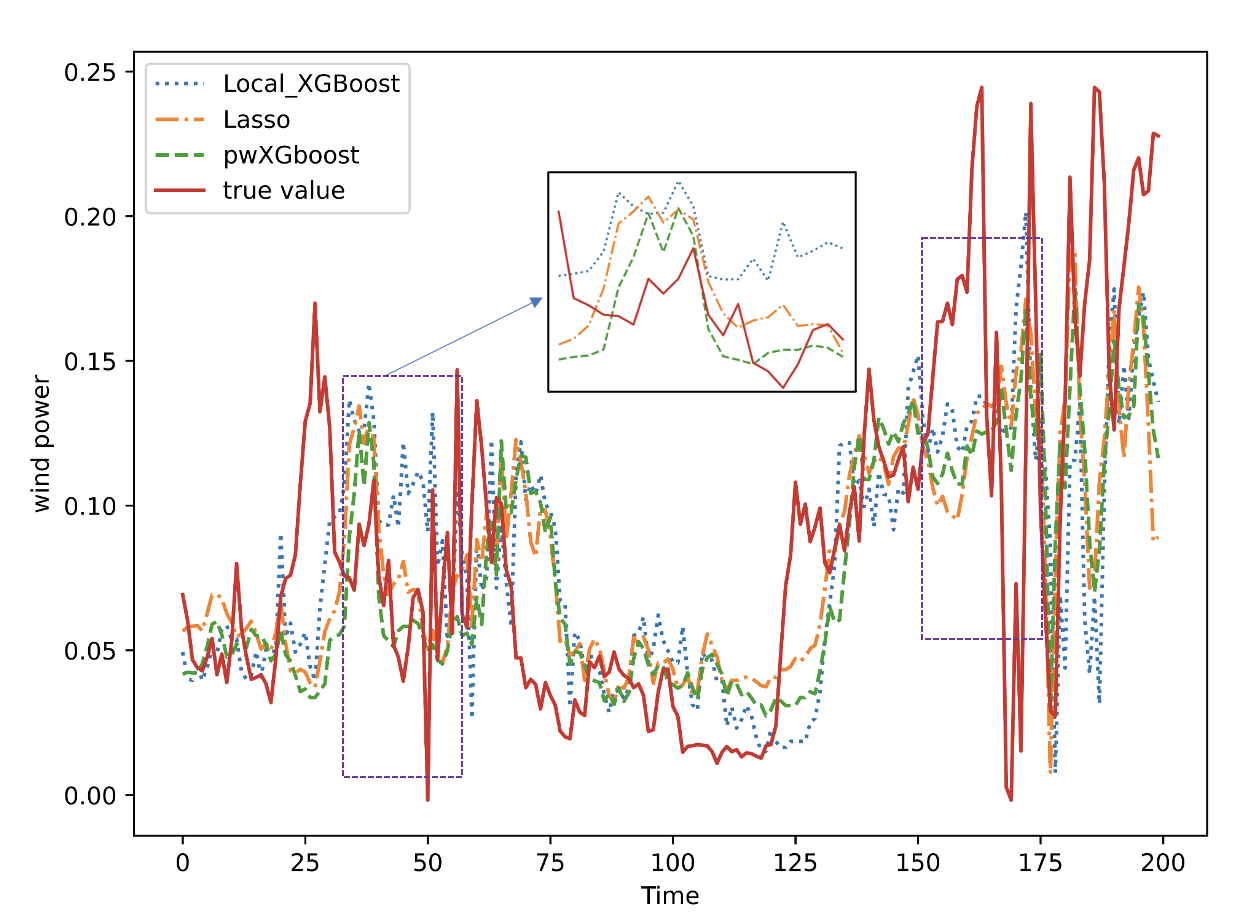}
    \caption{The Comparison of Different Prediction Results}
    \label{fig:prob}
\end{figure}

The prediction results of \emph{pwXGBoost}, \emph{Lasso} and \emph{Local\_XGBoost} are compared in Figure~\ref{fig:prob}. In this test, it is obvious that the wind power prediction method which uses the information from nearby wind farms performs better than the method only uses the local data. 

According to the prediction results in Figure~\ref{fig:prob}, we can also see that the pwXGBoost model is better than the linear method especially in scenarios with rapid changes in wind power which is crucial for the safe operation of power system. For the linear Lasso method, although it is easy to be implemented, its ability to capture the trend and extreme values is inferior to the nonlinear pwXGBoost.

\subsection{The Correlation Analysis of the Wind Farms}
\begin{figure}[htbp]
    \centering
    \includegraphics[width=0.8\textwidth]{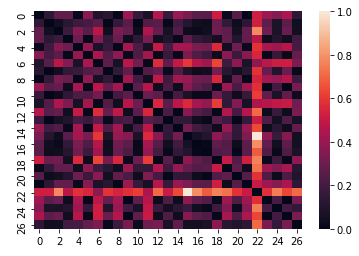}
    \caption{The Correlation of Wind Farms Based on MMD}
    \label{fig:corr}
\end{figure}

In this section, we demonstrate the effectiveness of the participant selection stage.
The MMD distances between each pair of wind farms are shown in Figure~\ref{fig:corr}.

\begin{figure}[htbp]
    \centering
    \includegraphics[width=1\textwidth]{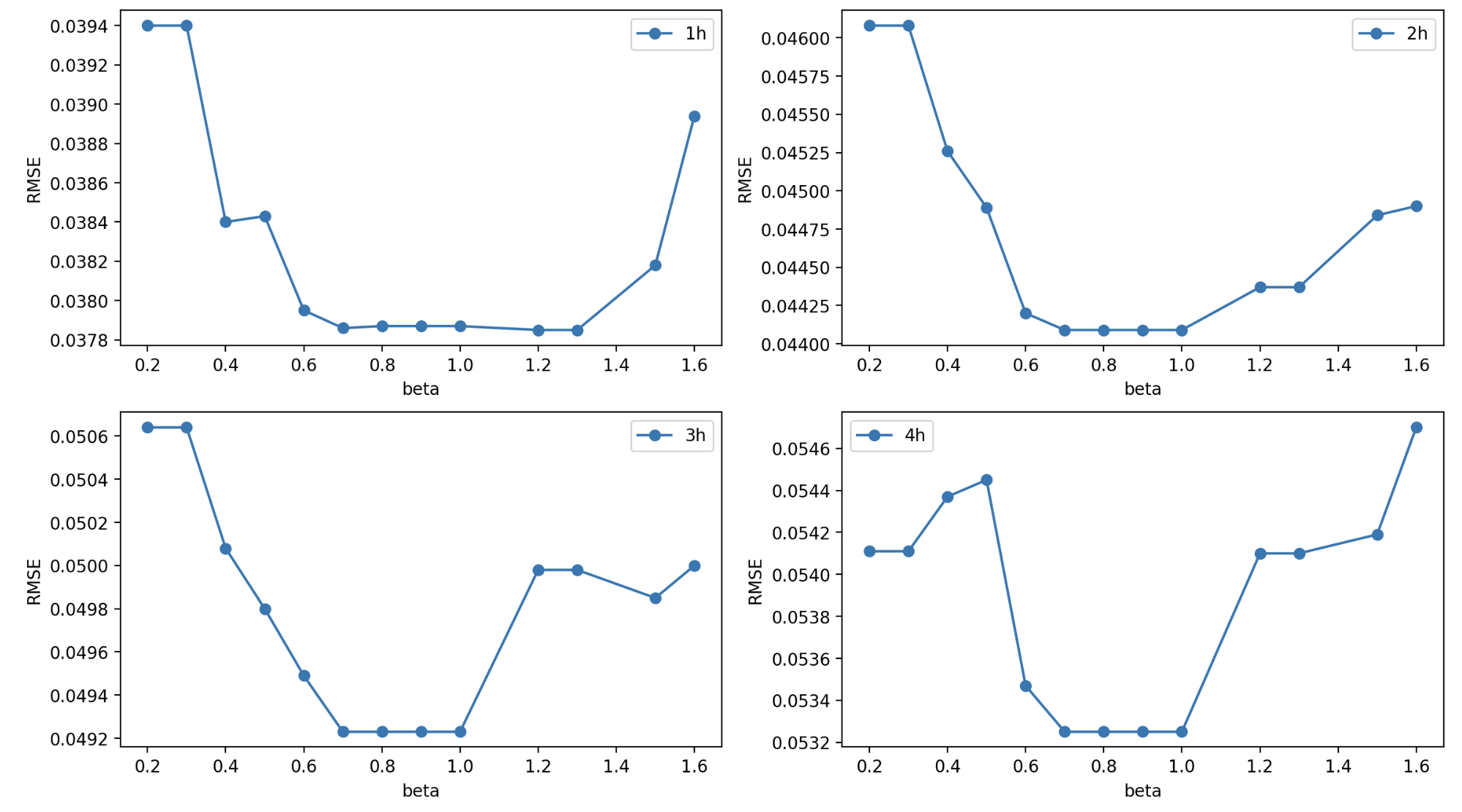}
    \caption{The RMSE Under Different MMD Threshold}
    \label{fig:rmse}
\end{figure}

We adjusted the value of the $\beta$ (the threshold of Equation~\ref{eq:11}), the wind farms can be divided into different groups and different number wind farms will be selected as the participant wind farms. Larger $\beta$ means more wind farms will be selected as the participant wind farms. The RMSE under different $\beta$ is shown in Figure~\ref{fig:rmse}.

The results show that when the threshold is in the range of 0.7 and 1, the number of selected wind farms are the same and the RMSE is the lowest.  
When the threshold is set too low ($< 0.7$), fewer wind farms will participate, and spatial and temporal correlation information is also constrained, resulting in an increase in the RMSE. On the other hand, if the threshold is set too high ($>1.0$), the various participating wind farms may introduce too diverse distributions, also increasing the RMSE.


\subsection{Scalability and Computation Cost}
The training time and prediction time of the privacy preserving ultra-short-term wind power prediction model is also important for the practical appplication. If it consumes a lot of time when training and prediction, it is not applicable and acceptable. Therefore, we test the consuming time of the privacy preserving XGBoost.\par

\begin{table}[htbp]
	\caption{Time Test of Training and Prediction Process}\label{tab:time}
    \begin{tabular}{|c|c|c|c|c|}\hline
    \multirow{2}*{Parties} & \multicolumn{2}{c|}{Training Time (s)}  & \multicolumn{2}{c|}{Inference time (s)} \\ 
    \cline{2-5}
    ~ & Linear Reg & XGB & Linear Reg & XGB \\ \hline 
    5 & 108.1 & 1421.0 &11.2 & 155.3\\ \hline
    10 & 111.3 & 2276.4 &11.3 & 284.9\\ \hline
    15 & 131.4 & 3181.0 &13.2 & 402.8\\ \hline
    20 & 140.6 & 3958.7 &14.2 & 566.5\\ \hline
    25 & 161.0 & 4736.8 &14.2 & 659.9\\ \hline
    27 & 169.1 & 5214.9 &15.2 & 714.3\\ \hline
\end{tabular}
\end{table}

From Table~\ref{tab:time}, it can be seen that even when all the 27 wind farms are included in the training and prediction process, the training time is less than 1.5h and the prediction time is less than 12min. In our prediction process, the 12 wind farms are selected based on the MMD method, the training time are reduced to less than 53 minutes and the prediction time are less than 400s, which is acceptable in practice.

\section{Conclusion} \label{sec:conclu}
Exploiting spatial and temporal correlation is useful to improve the accuracy of the ultra-short-term wind power. Taking data security and regulation policy issues into account, this paper proposes a privacy-preserving ultra-short-term wind power prediction method pwXGBoost based on secret sharing protocol and carefully chosen collaborative neighboring wind farms. The test results on the Inner Mongolian data set proved the ability of nonlinear feature extraction compared to the linear model and the data security is guaranteed mathematically. \par

According to the case study results, the RMSE of the method which uses the selected wind farms to train the prediction model is 0.4\% lower than that of the method which only uses the local wind farm data in the $4-th$ hour. In the meanwhile, the prediction time of privacy preserving method is less than 5min. Therefore, this method is acceptable for the application.\par

Besides, the secure multi party computation theory is scalable to the other nonlinear operation. Since neural network is another widely used method in wind farm power prediction, we can also build neural network based on secure multi party computation theory and secret share protocol. In the future, we will adopted the secret sharing protocol to the neural network and designed the privacy preserving wind power prediction method based on neural network to test its performance. 

\appendix
\section{Secure Multi Party Computation} \label{app:secmpc}

Secure Multi Party Computation (MPC) has a long history in the cryptography community,  it enables a group of data providers to jointly compute an agreed function without disclosing their data. One of the most fundamental building blocks of MPC is secret sharing, which is the basis of most current MPC platforms~\cite{keller2020mp, li2019privpy}.
$t$-out-of-$n$ secret sharing splits a secret input $x$ into $n$ shares, satisfying that any $t$ shares can completely reconstruct $x$ while any shares less than $t$ reveal nothing. For example, a commonly-used 2-out-of-3 secret sharing protocol splits $x$ into $[x] = \{[x]_1, [x]_2, [x]_3\} = \{(x_1, x_2), (x_2, x_3), (x_3, x_1)\}$ and let three computation servers hold the shares, respectively~\cite{keller2020mp, mohassel2018aby3}. $[x]$ satisfy that  $x \equiv \sum_{j=1}^{j=3}x_j(\mod M)$ where $M$ is a large integer, usually $2^k$, making $x_j, i \in {1, 2, 3}$ uniformly distributed in a ring of $\mathbb{Z}_{2^k}$. It is obvious that any two servers (or more) can together construct $x$ while each single server learns nothing. 

To make MPC general (i.e., to support arbitrary functions), researchers and engineers have proposed so-called \textit{general-purpose} MPC platforms, which offers a series of basic secure operations like secure addition, subtraction, multiplication, comparison which can be composed together to support complex functions like square-root and division and more advanced functions like machine-learning functions (e.g., secure principal component analysis~\cite{fan2021ppca}). All the secure basic operations are cryptographic protocols among the computation servers and preserve the privacy of the secret input $x$ during the computation. The computation process of these secret-sharing based MPC platforms has three stages:initialization stage, computation stage and reveal stage.

\paragraph{Secret share initialization stage.  } For $N$ data providers, they encode the original data $x^{(i)}, i \in {1, ..., N}$ into the secret share $[x^{(i)}]$ according to the specific $t$-out-of-$n$ secret sharing protocol and pass the each share to the target computation server.  In the end of this stage, each computation server $s_j, j \in {1, ..., n}$ will hold $N$ secret shares of the all the data providers, $\{[x^{(i)}]_j\}_{i=1}^{i=N}$.
In the above 2-out-of-3 secret sharing protocol, the three servers will hold $\{(x^{(i)}_{1}, x^{(i)}_{2})\}_{i=1}^{i=N}$, $\{(x^{(i)}_{2}, x^{(i)}_{3})\}_{i=1}^{i=N}$, $\{(x^{(i)}_{1}, x^{(i)}_{3})\}_{i=1}^{i=N}$, respectively. 

\paragraph{Computation stage.  } For any valid function $f(x^{i}, i \in {1, ..., n})$ that all the data providers agreed, it will be constructed as the \textit{modular composition} of the basic secure operations, which means that in the end of each secure operation, the computation servers will hold the secret shares of the corresponding result. For example, after the evaluation of secure addition between secret shares $[x]$ and $[y]$, the computation servers should hold the secret share $[z]$ where $z = x + y$. 
In this stage, all the $n$ computation servers will evaluate the composed secure operations sequentially for the final result. 

\paragraph{Reveal stage.  } After the computation stage of $f$, the computation servers will hold the secret shares of the final result (i.e., $[f(x^{i}, i \in {1, ..., n})]$). In this final stage, all the computation servers will pass its share to one party (predefined in the beginning) to reconstruct $f(x^{i}, i \in {1, ..., n})$. In fact, the reveal stage can based on the blockchain technology to regulate the information exchange and its profit allocation \cite{cui2021blockchain}.\par

\section{Secure Operations for Secret Sharing Based MPC Platforms.}

In this section, we first introduce some basic secure operations briefly, showing how the secret sharing computation works. In the end, we introduce secure division operation to demonstrate how to compose basic operations for complex functions. 

\paragraph{Secure addition.  } For two data $x$ and $y$, their secret shares are $[x]$ and $[y]$. It is self-evidently additive homomorphic because $[x] + [y] = [x + y]$. Each computation server can locally compute the share of the sum in the computation process.

\paragraph{Secure subtraction.  } In the secret sharing computation, addition and subtraction are equivalent, as $x$ minus $y$ is equivalent to $x$ plus the opposite of $y$. Servers can firstly construct the share of $[-y]$ locally by computing the opposite of its own shares as $ -y \equiv \sum_{i=1}^{i=n} - y_i(\mod M)$. Then compute $[x - y]$ through $[x] + [-y]$.


\paragraph{Secure multiplication.  } Computing multiplication usually requires communications among the computation servers. 
Take the above 2-out-of-3 secret sharing and ABY3~\cite{mohassel2018aby3} multiplication protocol for example, defining $z = xy$.  As $x = (x_1 + x_2 + x_3)(y_1 + y_2 + y_3)$, we can define $[z]$ as $z_1 = x_1y_1 + x_1y_2 + x_2y_1 + \alpha_1$, $z_2 = x_2y_2 + x_2y_3 + x_3y_2 + \alpha_2$ and $z_3 = x_3y_3 + x_3y_1 + x_1y_3 + \alpha_3$ where $\sum_{j=1}^{j=3}{\alpha_j} = 0$. Each computation server $j$ firstly locally compute its share $z_j$, then communicate its share to the previous server for valid secret shares (as we require each server hold two shares from $\{z_1, z_2, z_3\}$).



\paragraph{Secure comparison.  } When two numbers $x$ and $y$ are compared, $x < y$ equals a Boolean indicating whether $(x-y)$ is negative. So the computation servers can firstly compute $[x-y]$ based on the secure subtraction protocol and then use the \textit{bit extraction protocol}~\cite{nishide2007multiparty} to securely extract the sign bit of $[x-y]$. If the highest bit is 1, it means that $x-y$ is negative and therefore $x<y$.\par

\paragraph{Secure division.  } Comparing with other basic operations, computing non-linear functions like secure division is more involved. As $[x]$ divides $[y]$ is equivalent to $[x]$ multiply $[\frac{1}{y}]$, how to compute the reciprocal of $[y]$ (i.e., $[\frac{1}{y}]$) is important. 

As $\frac{1}{y}$ is the solution of Equation~\ref{eq:div1}, it is obvious that the solution $x = \frac{1}{y}$. 
\begin{equation}\label{eq:div1}
\begin{aligned}
    f(x) = \frac{1}{x}-y = 0
\end{aligned}
\end{equation}

As we know, the Taylor expansion of $f(x)$ is as follows:
\begin{equation}\label{eq:div2}
\begin{aligned}
f(x) = f(x_0)+\frac{f^{'}(x_0)}{1!}(x-x_0)+\frac{f^{''}(x_0)}{2!}(x-x_0)^{2}+ 
        ...+\frac{f^{n}(x_0)}{n!}(x-x_0)^{n}
\end{aligned}
\end{equation}

The first order approximation of $f(x)$ is used to get the solution which is the Newton-Raphson method. The approximate solution can be calculated as follows:
\begin{equation}
\begin{aligned}
x_{n+1} = x_{n}-\frac{f(x_n)}{f^{'}(x_{n})}=2x_{n}-ax^{2}_{n}
\end{aligned}
\end{equation}

Thus, we can compose a series of secure subtraction and multiplication to update $x_{n+1}$ till it converges to the expected reciprocal. 
\bibliographystyle{unsrt}
\bibliography{cas-refs}

\begin{thebibliography}{10}

\bibitem{he2016cooperation}
Guannan He, Qixin Chen, Chongqing Kang, Qing Xia, and Kameshwar Poolla.
\newblock Cooperation of wind power and battery storage to provide frequency
  regulation in power markets.
\newblock {\em IEEE Transactions on Power Systems}, 32(5):3559--3568, 2016.

\bibitem{fabbri2005assessment}
Alberto Fabbri, T~GomezSan Roman, J~Rivier Abbad, and VH~M{\'e}ndez Quezada.
\newblock Assessment of the cost associated with wind generation prediction
  errors in a liberalized electricity market.
\newblock {\em IEEE Transactions on Power Systems}, 20(3):1440--1446, 2005.

\bibitem{tastu2013probabilistic}
Julija Tastu, Pierre Pinson, Pierre-Julien Trombe, and Henrik Madsen.
\newblock Probabilistic forecasts of wind power generation accounting for
  geographically dispersed information.
\newblock {\em IEEE Transactions on Smart Grid}, 5(1):480--489, 2013.

\bibitem{fan2020m2gsnet}
Hang Fan, Xuemin Zhang, Shengwei Mei, Kunjin Chen, and Xinyang Chen.
\newblock {M2GSN}et: Multi-modal multi-task graph spatiotemporal network for
  ultra-short-term wind farm cluster power prediction.
\newblock {\em Applied Sciences}, 10(21):7915, 2020.

\bibitem{gonccalves2021critical}
Carla Gon{\c{c}}alves, Ricardo~J Bessa, and Pierre Pinson.
\newblock A critical overview of privacy-preserving approaches for
  collaborative forecasting.
\newblock {\em International journal of Forecasting}, 37(1):322--342, 2021.

\bibitem{dwork2008differential}
Cynthia Dwork.
\newblock Differential privacy: A survey of results.
\newblock In {\em International conference on theory and applications of models
  of computation}, pages 1--19. Springer, 2008.

\bibitem{abadi2016deep}
Martin Abadi, Andy Chu, Ian Goodfellow, H~Brendan McMahan, Ilya Mironov, Kunal
  Talwar, and Li~Zhang.
\newblock Deep learning with differential privacy.
\newblock In {\em Proceedings of the 2016 ACM SIGSAC conference on computer and
  communications security}, pages 308--318, 2016.

\bibitem{goncalves2021privacy}
Carla Goncalves, Ricardo~J Bessa, and Pierre Pinson.
\newblock Privacy-preserving distributed learning for renewable energy
  forecasting.
\newblock {\em IEEE Transactions on Sustainable Energy}, 12(3):1777--1787,
  2021.

\bibitem{goncalves2020towards}
Carla Goncalves, Pierre Pinson, and Ricardo~J Bessa.
\newblock Towards data markets in renewable energy forecasting.
\newblock {\em IEEE Transactions on Sustainable Energy}, 12(1):533--542, 2020.

\bibitem{han2021trading}
Liyang Han, Pierre Pinson, and Jalal Kazempour.
\newblock Trading data for wind power forecasting: A regression market with
  lasso regularization.
\newblock {\em arXiv preprint arXiv:2110.07432}, 2021.

\bibitem{kairouz2021advances}
Peter Kairouz, H~Brendan McMahan, Brendan Avent, Aur{\'e}lien Bellet, Mehdi
  Bennis, Arjun~Nitin Bhagoji, Kallista Bonawitz, Zachary Charles, Graham
  Cormode, Rachel Cummings, et~al.
\newblock Advances and open problems in federated learning.
\newblock {\em Foundations and Trends{\textregistered} in Machine Learning},
  14(1--2):1--210, 2021.

\bibitem{yao1982protocols}
Andrew~C Yao.
\newblock Protocols for secure computations.
\newblock In {\em 23rd annual symposium on foundations of computer science
  (sfcs 1982)}, pages 160--164. IEEE, 1982.

\bibitem{yang2019federated}
Qiang Yang, Yang Liu, Tianjian Chen, and Yongxin Tong.
\newblock Federated machine learning: Concept and applications.
\newblock {\em ACM Transactions on Intelligent Systems and Technology (TIST)},
  10(2):1--19, 2019.

\bibitem{ogburn2013homomorphic}
Monique Ogburn, Claude Turner, and Pushkar Dahal.
\newblock Homomorphic encryption.
\newblock {\em Procedia Computer Science}, 20:502--509, 2013.

\bibitem{yao1986generate}
Andrew Chi-Chih Yao.
\newblock How to generate and exchange secrets.
\newblock In {\em 27th Annual Symposium on Foundations of Computer Science
  (sfcs 1986)}, pages 162--167. IEEE, 1986.

\bibitem{bellare2012foundations}
Mihir Bellare, Viet~Tung Hoang, and Phillip Rogaway.
\newblock Foundations of garbled circuits.
\newblock In {\em Proceedings of the 2012 ACM conference on Computer and
  communications security}, pages 784--796, 2012.

\bibitem{rabin2005exchange}
Michael~O Rabin.
\newblock How to exchange secrets with oblivious transfer.
\newblock {\em Cryptology ePrint Archive}, 2005.

\bibitem{shamir1979share}
Adi Shamir.
\newblock How to share a secret.
\newblock {\em Communications of the ACM}, 22(11):612--613, 1979.

\bibitem{li2019privpy}
Yi~Li and Wei Xu.
\newblock {P}riv{P}y: General and scalable privacy-preserving data mining.
\newblock In {\em Proceedings of the 25th ACM SIGKDD International Conference
  on Knowledge Discovery \& Data Mining}, pages 1299--1307, 2019.

\bibitem{keller2020mp}
Marcel Keller.
\newblock {MP-SPDZ}: A versatile framework for multi-party computation.
\newblock In {\em Proceedings of the 2020 ACM SIGSAC conference on computer and
  communications security}, pages 1575--1590, 2020.

\bibitem{tian2021fully}
Nianfeng Tian, Qinglai Guo, Hongbin Sun, and Xin Zhou.
\newblock Fully privacy-preserving distributed optimization based on secret
  sharing.
\newblock 2021.

\bibitem{li2020federated}
Tian Li, Anit~Kumar Sahu, Ameet Talwalkar, and Virginia Smith.
\newblock Federated learning: Challenges, methods, and future directions.
\newblock {\em IEEE Signal Processing Magazine}, 37(3):50--60, 2020.

\bibitem{liu2020secure}
Yang Liu, Yan Kang, Chaoping Xing, Tianjian Chen, and Qiang Yang.
\newblock A secure federated transfer learning framework.
\newblock {\em IEEE Intelligent Systems}, 35(4):70--82, 2020.

\bibitem{toubeau2022privacy}
Jean-Fran{\c{c}}ois Toubeau, Fei Teng, Thomas Morstyn, Leandro
  Von~Krannichfeldt, and Yi~Wang.
\newblock Privacy-preserving probabilistic voltage forecasting in local energy
  communities.
\newblock {\em IEEE Transactions on Smart Grid}, 14(1):798--809, 2022.

\bibitem{wang2023federated}
Yi~Wang, Jiahao Ma, Ning Gao, Qingsong Wen, Liang Sun, and Hongye Guo.
\newblock Federated fuzzy k-means for privacy-preserving behavior analysis in
  smart grids.
\newblock {\em Applied Energy}, 331:120396, 2023.

\bibitem{qiu2023federated}
Dawei Qiu, Juxing Xue, Tingqi Zhang, Jianhong Wang, and Mingyang Sun.
\newblock Federated reinforcement learning for smart building joint
  peer-to-peer energy and carbon allowance trading.
\newblock {\em Applied Energy}, 333:120526, 2023.

\bibitem{li2023wind}
Yang Li, Ruinong Wang, Yuanzheng Li, Meng Zhang, and Chao Long.
\newblock Wind power forecasting considering data privacy protection: A
  federated deep reinforcement learning approach.
\newblock {\em Applied Energy}, 329:120291, 2023.

\bibitem{sun2019can}
Ziteng Sun, Peter Kairouz, Ananda~Theertha Suresh, and H~Brendan McMahan.
\newblock Can you really backdoor federated learning?
\newblock {\em arXiv preprint arXiv:1911.07963}, 2019.

\bibitem{acar2018survey}
Abbas Acar, Hidayet Aksu, A~Selcuk Uluagac, and Mauro Conti.
\newblock A survey on homomorphic encryption schemes: Theory and
  implementation.
\newblock {\em ACM Computing Surveys (Csur)}, 51(4):1--35, 2018.

\bibitem{li2021federated}
Qinbin Li, Yiqun Diao, Quan Chen, and Bingsheng He.
\newblock Federated learning on non-{IID} data silos: An experimental study.
\newblock {\em arXiv preprint arXiv:2102.02079}, 2021.

\bibitem{fontaine2007survey}
Caroline Fontaine and Fabien Galand.
\newblock A survey of homomorphic encryption for nonspecialists.
\newblock {\em EURASIP Journal on Information Security}, 2007:1--10, 2007.

\bibitem{gouert2022new}
Charles Gouert, Dimitris Mouris, and Nektarios~Georgios Tsoutsos.
\newblock New insights into fully homomorphic encryption libraries via
  standardized benchmarks.
\newblock {\em Cryptology ePrint Archive}, 2022.

\bibitem{wang2012accelerating}
Wei Wang, Yin Hu, Lianmu Chen, Xinming Huang, and Berk Sunar.
\newblock Accelerating fully homomorphic encryption using {GPU}.
\newblock In {\em 2012 IEEE conference on high performance extreme computing},
  pages 1--5. IEEE, 2012.

\bibitem{tan2022towards}
Alysa~Ziying Tan, Han Yu, Lizhen Cui, and Qiang Yang.
\newblock Towards personalized federated learning.
\newblock {\em IEEE Transactions on Neural Networks and Learning Systems},
  2022.

\bibitem{chen2016xgboost}
Tianqi Chen and Carlos Guestrin.
\newblock {XGBoost}: A scalable tree boosting system.
\newblock In {\em Proceedings of the 22nd acm sigkdd international conference
  on knowledge discovery and data mining}, pages 785--794, 2016.

\bibitem{nishide2007multiparty}
Takashi Nishide and Kazuo Ohta.
\newblock Multiparty computation for interval, equality, and comparison without
  bit-decomposition protocol.
\newblock In {\em International Workshop on Public Key Cryptography}. Springer,
  2007.

\bibitem{zhuang2020comprehensive}
Fuzhen Zhuang, Zhiyuan Qi, Keyu Duan, Dongbo Xi, Yongchun Zhu, Hengshu Zhu, Hui
  Xiong, and Qing He.
\newblock A comprehensive survey on transfer learning.
\newblock {\em Proceedings of the IEEE}, 109(1):43--76, 2020.

\bibitem{zheng2019xgboost}
Huan Zheng and Yanghui Wu.
\newblock A xgboost model with weather similarity analysis and feature
  engineering for short-term wind power forecasting.
\newblock {\em Applied Sciences}, 9(15):3019, 2019.

\bibitem{li2020short}
Wenze Li, Xiaosheng Peng, Kai Cheng, Hongyu Wang, Qiyou Xu, Bo~Wang, and
  Jianfeng Che.
\newblock A short-term regional wind power prediction method based on xgboost
  and multi-stage features selection.
\newblock In {\em 2020 IEEE 3rd Student Conference on Electrical Machines and
  Systems (SCEMS)}, pages 614--618. IEEE, 2020.

\bibitem{ke2017lightgbm}
Guolin Ke, Qi~Meng, Thomas Finley, Taifeng Wang, Wei Chen, Weidong Ma, Qiwei
  Ye, and Tie-Yan Liu.
\newblock {L}ight{GBM}: A highly efficient gradient boosting decision tree.
\newblock {\em Advances in neural information processing systems}, 30, 2017.

\bibitem{prokhorenkova2018catboost}
Liudmila Prokhorenkova, Gleb Gusev, Aleksandr Vorobev, Anna~Veronika Dorogush,
  and Andrey Gulin.
\newblock {C}at{B}oost: unbiased boosting with categorical features.
\newblock {\em Advances in neural information processing systems}, 31, 2018.

\bibitem{ziller2021pysyft}
Alexander Ziller, Andrew Trask, Antonio Lopardo, Benjamin Szymkow, Bobby
  Wagner, Emma Bluemke, Jean-Mickael Nounahon, Jonathan Passerat-Palmbach,
  Kritika Prakash, Nick Rose, et~al.
\newblock {P}y{S}yft: A library for easy federated learning.
\newblock In {\em Federated Learning Systems}, pages 111--139. Springer, 2021.

\bibitem{he2020fedml}
Chaoyang He, Songze Li, Jinhyun So, Xiao Zeng, Mi~Zhang, Hongyi Wang, Xiaoyang
  Wang, Praneeth Vepakomma, Abhishek Singh, Hang Qiu, et~al.
\newblock {F}ed{ML}: A research library and benchmark for federated machine
  learning.
\newblock {\em arXiv preprint arXiv:2007.13518}, 2020.

\bibitem{liu2021fate}
Yang Liu, Tao Fan, Tianjian Chen, Qian Xu, and Qiang Yang.
\newblock {FATE}: An industrial grade platform for collaborative learning with
  data protection.
\newblock {\em J. Mach. Learn. Res.}, 22(226):1--6, 2021.

\bibitem{ben2016optimizing}
Aner Ben-Efraim, Yehuda Lindell, and Eran Omri.
\newblock Optimizing semi-honest secure multiparty computation for the
  {I}nternet.
\newblock In {\em Proceedings of the ACM SIGSAC Conference on Computer and
  Communications Security (CCS)}, 2016.

\bibitem{patel2015efficient}
Sankita~J Patel, Dharmen Punjani, and Devesh~C Jinwala.
\newblock An efficient approach for privacy preserving distributed clustering
  in semi-honest model using elliptic curve cryptography.
\newblock {\em International Journal of Network Security}, 2015.

\bibitem{fan2021ppca}
Xiaoyu Fan, Guosai Wang, Kun Chen, Xu~He, and Wei Xu.
\newblock {PPCA}: Privacy-preserving principal component analysis using secure
  multiparty computation ({MPC}).
\newblock {\em arXiv preprint arXiv:2105.07612}, 2021.

\bibitem{evans2018pragmatic}
David Evans, Vladimir Kolesnikov, Mike Rosulek, et~al.
\newblock A pragmatic introduction to secure multi-party computation.
\newblock {\em Foundations and Trends{\textregistered} in Privacy and
  Security}, 2018.

\bibitem{canetti2000security}
Ran Canetti.
\newblock Security and composition of multiparty cryptographic protocols.
\newblock {\em Journal of CRYPTOLOGY}, 2000.

\bibitem{burns2022kubernetes}
Brendan Burns, Joe Beda, Kelsey Hightower, and Lachlan Evenson.
\newblock {\em Kubernetes: up and running}.
\newblock " O'Reilly Media, Inc.", 2022.

\bibitem{mohassel2018aby3}
Payman Mohassel and Peter Rindal.
\newblock Aby3: A mixed protocol framework for machine learning.
\newblock In {\em Proceedings of the ACM SIGSAC conference on computer and
  communications security (CCS)}, 2018.

\bibitem{cui2021blockchain}
Jingshi Cui, Nan Gu, and Chenye Wu.
\newblock Blockchain enabled data transmission for energy imbalance market.
\newblock {\em IEEE Transactions on Sustainable Energy}, 13(2):1254--1266,
  2021.

\end{thebibliography}


\end{document}